\documentclass[aps,prd,amsmath,amssymb,preprintnumbers,nofootinbib,preprint]{revtex4-1}
\usepackage{amsthm}
\usepackage{color}
\usepackage{dcolumn}% Align table columns on decimal point
\usepackage{bm}% bold math
\usepackage{bbm}
\usepackage{pxfonts}
\usepackage{slashed} 
\usepackage{url}
\usepackage{axodraw}
\usepackage{graphicx}

\usepackage[ margin=5pt, font=normalsize,labelfont=bf,justification=raggedright]{caption}

\def\ga{\gamma}

\def\la{\lambda}

\def\Ga{\Gamma}

\def\pa{\partial}

\newcommand{\rL}{\rho_{\Lambda}}

\newcommand{\rLV}{\rL^{\rm vac}}

\newcommand{\MS}{{\mbox{\sc mom}}}

\newcommand{\be}{\begin{eqnarray}}
\newcommand{\ee}{\end{eqnarray}}

\begin{document}

%%%%%%%%%%%%%%%%%%%%%%%%%%%%%%%%%%%%%%%%%%%%%%%%%%%%%%%%%%%%%%%%%%%%%%%%%%%
\title{Revisiting the decoupling effects in the running of the Cosmological Constant }
%%%%%%%%%
\author{Oleg Antipin}%$^{\color{rossoCP3}{\varheartsuit}}$}
\email{oantipin@irb.hr}
\author{Bla\v zenka Meli\'c \vspace{4mm}}
 \email{melic@irb.hr}
\affiliation{Rudjer Bo\v{s}kovi\'c Institute, Division of Theoretical Physics,
Bijeni\v{c}ka 54, HR-10000 Zagreb, Croatia\\}
%%%%%
%%%%%%%%%%%%%%%%%%%%%%%%%%%%%%%%%%%%%%%%%%%%%%%%%%%%%%%%%%%%%%%%%%%%%%%%%%%
\begin{abstract}
\vspace{4mm}
  We revisit the decoupling effects associated with heavy particles in the renormalization group running of the vacuum energy in a mass-dependent renormalization scheme. We find the running of the vacuum energy stemming from the Higgs condensate in the entire energy range and show that it behaves as expected from the simple dimensional arguments meaning that it exhibits the quadratic sensitivity to the mass of the heavy particles in the infrared regime. The consequence of such a running to the fine-tuning problem with the measured value of the Cosmological Constant is analyzed 
and the constraint on the mass spectrum of a given model is derived. We show that in the Standard Model (SM) this fine-tuning constraint is not satisfied while in the massless theories this constraint formally coincides with the well known Veltman condition. We also provide a remarkably simple extension of the SM where saturation of this constraint enables us to predict the radiative Higgs mass correctly. Generalization to constant curvature spaces is also given.
\end{abstract}
%  Such an enhanced running would lead to severe fine-tuning problem with the measured value of the %Cosmological Constant unless we constrain the heavy masses spectrum to cancel this leading effect in the beta %function. 

%\arxivnumber{1703.10967}  
%  and show that it has 
 % while the running of the vacuum part of the CC cannot be accessed within our framework. Moreover, we show that in the massless theories
 %
%\\[.5cm]
%{\footnotesize  \it Preprint: }
%\end{abstract}  

%\begin{document}
\maketitle

\section{Introduction}
  
It is widely accepted that our today's universe is undergoing the phase of the accelerated expansion which is commonly explained by the presence of the Cosmological Constant (CC) $\Lambda$. However, the value of $\Lambda$ required by experiment is in a contradiction with the values emerging from the physics scales associated with known phase transitions in the universe so that severe fine-tuning has to be applied which is at heart of the CC problem. 
To recall the main aspects of this problem we begin with the Standard Model (SM) formulated on the classical curved background. In order to
construct a renormalizable gauge theory in an external gravitational field
one starts from the classical action (with $\varphi$ as the Higgs doublet field)
\begin{equation}
S_{vac}=\int d^{4}x\sqrt{-g}\,\Big\{ \mathcal{L}_{SM}+ \,  \xi \,\,\varphi ^{\dagger }\varphi \,R+ a_{1}R^{2}_{\mu\nu\alpha\beta}
+a_{2}R^{2}_{\mu\nu} + a_{3}R^{2} + a_{4}{\square} R-\frac{1}{16\pi G_{vac}}%
\,(R + 2\Lambda_{vac})\Big\} \,. \label{1}
\end{equation}
The renormalization procedure for the theory (\ref{1}) consists of the renormalization of the SM matter fields, couplings and masses, non-minimal
coupling $\xi$ and the gravitational couplings $a_{1,2,3,4},$ 
$\,G_{vac}\,$ and $\,\Lambda_{vac}$. We are going to work in the low energy domain of the gravitational physics
and, for that reason, the short distance effects from the higher
derivative terms $a_{1,2,3,4},$ in (\ref{1}) are not important for our considerations, and
so we start with the usual bare Hilbert-Einstein action 
%with a running cosmological and gravitational
with coupling constants $G_{vac},\,\Lambda _{vac}$ supplemented with non-minimal coupling $\xi$:
\begin{equation}
S_{HE}=\int d^{4}x\sqrt{-g}\,\left\{ \,%
\,\mathcal{L}_{SM}+  \xi^0 \,\,\varphi^{\dagger }\varphi \, R -\frac{1}{16\pi G_{vac}^0}(R +2\Lambda _{vac}^0)\right\} \,.  \label{HE}
\end{equation}
The bare quantities are defined with the superscript "0". 
Let us focus on the CC itself which, as we mentioned above,
must be renormalized and the connection with experimentally measured value $\rho_{phys}$ is achieved via the
renormalization condition (see Eq.(\ref{CCphys}) below) imposed on the vacuum energy density:
\be (\rho^0_{\Lambda})^{vac} = \frac{\Lambda^0_{vac}}{8\pi G^0_{vac}} \, \label{CCd3} \ee
%The numerical value of $\Lambda _{vac}$ should resultfrom the experiment.
at some energy scale $\mu$ so that $\rLV(\mu)$.
Moreover,  in the presence of the dynamical cosmological background characterized by the time-dependent Hubble parameter $H(t)$, $\rLV$ can be dynamical $\rLV(\mu, t)$
that will be reflected in the evolution of $\rLV$ via:
\be
\frac{\pa \rLV(t)}{dt} = F(H, \rho_{\rm matter}, \rLV,....) \,,
\label{timeCC}
\ee
which is the important ingredient for cosmological evolution. Currently, there is no consensus on whether $\rLV(\mu, t)$ depends on $t$. 
Even if it is time-dependent, to understand the precise form of (\ref{timeCC}) one may go back to the \emph{time-independent} RG problem:
\be
\frac{\pa \rLV(\mu)}{d\mu} = \beta_{\rLV}(g_i, m_i, \mu,...) \,,
\label{muCC}
\ee
where $g_i$ and $m_i$ are the dimensionless couplings and masses respectively. The $g_i$ and $m_i$
are also supplemented with their own Renormalization Group (RG) equations.

Besides $\rLV$, the physical vacuum energy $\rho_{phys}$ consists of several additional parts.
% However, there is an essential difference between the
%CC and, say, masses of the particles. The key point of the CC problem is
%that, along with the $\rLV(\mu)$, there are other contributions to the physical ($\mu$-independent) vacuum energy $\rho_{phys}$. 
One of these parts is ''induced'' contribution $\rho _{ind}(\mu)\,$ to
the vacuum energy density arising from the vacuum condensates. For example, if $\varphi_{vac}$ is the value of the Higgs field $\varphi (x)$ which minimizes the 
Higgs potential $V(\varphi )$
%, then the lowest state has $T_{\mu\nu} = g_{\mu\nu}
%V(\varphi_{vac})$, which is the classical scalar field contribution to the 
%vacuum energy. Concretely, minimizing the Higgs
%potential 
\be
V(\varphi ) = -m^2 \varphi^{\dagger}\varphi +\frac{\lambda}{2}(\varphi^{\dagger}\varphi )^2 \,,
\label{higgspotential}
\ee 
the Higgs condensate contribution (at 
the classical level) to the vacuum energy is
\begin{equation}
\label{eq:cond}
\rho_{ind}(\mu) =V(\varphi_{vac} ) =-\frac{m^4(\mu)}{2 \lambda(\mu)} \, .
\end{equation}
Besides the
vacuum and induced terms we may have additional effects from the higher derivative terms in (\ref{1}) 
as well as corrections from quantum gravity. Again, these contributions can be classified as coming from
purely quantum effects and therefore expected to be $\mu$-dependent and some also time-dependent due to the expanding background, and therefore contributing to (\ref{timeCC}).
All in all, the physical value is measured at the cosmological RG scale $\mu _{c}$, which is experimentally given by $\mu _{c}=\mathcal{O}(10^{-3})\,$ eV, as
\begin{equation}
\rho _{phys}=\rLV(\mu_c)+\rho _{ind}(\mu_c)+ ...\,  =10^{-47} {\rm \ GeV}^4 \ .
\label{CCphys}
\end{equation}
The problem now is that if we use the experimental Higgs mass $\,M_{H}= 125\,$GeV, then the corresponding value $\left| \rho _{ind}\right| \simeq
10^{8}\,{\rm GeV}^{4}$. In order to keep the QFT consistent with astronomical
observations, one has to demand that the parts contributing to the $\rho_{phys}$ should cancel with the
accuracy dictated by the current data.  
%As shown by Eq.(\ref{eq:cond}), the term $\Lambda _{ind}$
%on the \textit{r.h.s.} of (\ref{lambda_phys}) is not an independent
%parameter of the SM, since it is constructed from other $m$ and $\lambda$ parameters. On the contrary, $\Lambda _{vac}$ is an
%independent parameter and requires an independent renormalization condition.
%From the QFT point of view, the sequence of steps in
%defining the CC is the following: one has to calculate the value of $\Lambda
%_{ind}$ at $\mu _{c}$, measure the value of the physical CC, $\Lambda _{phys}$%
%, at the same scale, and choose the renormalization condition for $\Lambda
%_{vac}$ in the form
%\begin{equation}
%\Lambda _{vac}({\mu _{c})}=\Lambda _{phys}({\mu _{c}})-\Lambda _{ind}({\mu _{c}%
%})\,.  \label{ren_cond}
%\end{equation}
%The modern observations tell us that
%the value of $\Lambda _{phys}(\mu =\mu _{c})$ is positive and has the
%magnitude of the order of $\rho _{c}^{0}$ , that is about $10^{-47}\,GeV^{4}$.
%Now, if we use $\,M_{H}= 125\,GeV$ then the corresponding value $\left| \Lambda _{ind}\right| \simeq
%1.0\times 10^{8}\,GeV^{4}$. These $\Lambda _{phys}$ and $\Lambda _{ind}$ values should be inserted into the RG condition (\ref
%{ren_cond}) and the problem now
For example, if we neglect all the $...$ terms in (\ref{CCphys}), the $\rLV$ and $\rho_{ind}$ should cancel
 with the precision of 55 decimal orders. This is the CC fine-tuning problem \cite{Shapiro:2009dh,Shapiro:2008sf}.

To understand deeper this tuning, one has to take into account the decoupling effects due to massive particles. Clearly, we expect
that contribution to the RG running from the particle of mass $m$ should change dramatically, as we go from $\mu \gg m$
to $\mu\ll m$ regime. Moreover, requiring the absence (or, at least, reduction) of the tuning may provide a constraint on the spectrum
of the particle physics models. In this paper, we deal with the time-independent classical curved background and will derive the RG evolution of  $\rLV$ and $\rho _{ind}$
of the form (\ref{muCC}) taking into account the decoupling effects due to massive particles by using the mass-dependent RG formalism. 
\\
\noindent

The motivation for this work is threefold:
\begin{itemize}
\item to derive the leading decoupling effects on the RG running of $\rho _{phys}$. Along the way, we will comment on the inconsistencies of the similar derivation presented in \cite{Babic:2001vv}, as we demonstrate the importance of considering the RG running of the total vacuum energy $\rLV+\rho _{ind}$ since, although $\rLV$ and $\rho_{ind}$ run separately, it is only the sum that exhibits behavior consistent with the Appelquist-Carazzone decoupling theorem \cite{Appelquist:1974tg}.  % as the contribution due to the Goldstone bosons comes with the opposite sign in two terms
%and cancels in the final sum. 
\item to elucidate the implications of these results on the mass spectrum of the SM as well as its extensions. As an outcome, we present a simple phenomenological extension of the SM \emph{predicting} the Higgs mass correctly.

\item to provide the generalization of these results to the constant curvature spaces important for studies of curvature-induced running of the vacuum energy as well as curvature-induced phase transitions.
\end{itemize}

The paper is structured as follows. In the next Sec.\ref{RGphi4} we briefly discuss the RG running of the CC in the simple $\phi^4$-theory highlighting
the necessary RG formalism we use later and also discuss the basic issue of decoupling in the RG running. In Sec.\ref{SMRG} we 
extend the RG approach to the full SM, in both, mass-independent and mass-dependent RG schemes. Sec.\ref{Appl} deals with applications of the derived heavy-mass threshold effects within and beyond the SM 
and Sec.\ref{Concl} presents our conclusions. In Appendices we provide the technical details, as well as generalize the flat spacetime results to the
spaces with constant curvature.
 
\section{RG Running of the Cosmological Constant }\label{RGphi4}

\begin{figure}[h]
 \center
\includegraphics[width=0.85\textwidth]{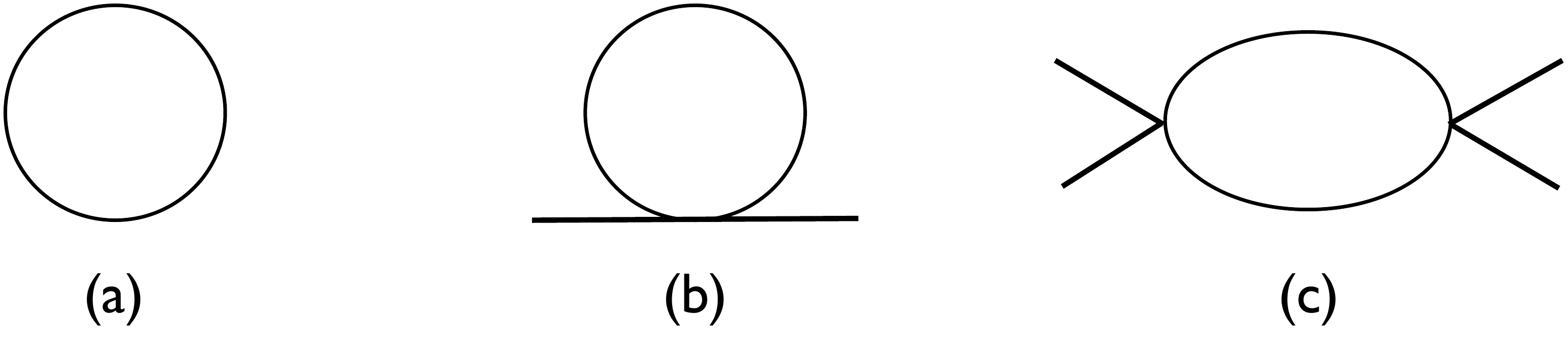}
\caption{\em
(\textit{a}) The one-loop contributions to the vacuum part of CC are just bubbles of matter fields without external legs; (\textit{b}) The one-loop two-point
function contributing to the induced part of the CC. (\textit{c}) The
one-loop four-point function contributing to the induced part of the CC.
\label{CC}}
\end{figure}

%The cosmological constant contribution is set by the value of the effective potential
%at its minimum. So, to 
To prepare for the discussion of the RG dependence of the CC and to setup the formalism, let us consider the skeleton Lagrangian for the real scalar:
\be
L=\frac{1}{2}m^2 \phi^2 +\frac{1}{8}\lambda \phi^4 \ .
\ee
The schematic contributions to the one-loop effective potential, up to the 4 external legs, are shown in Fig.\ref{CC} . 
%within the sharp cut-off approximation. 

Correspondingly to this diagramatic picture and for a general QFT, the renormalized effective potential can be split into two pieces: the $\phi$-independent (vacuum) term corresponding to the diagram Fig.\ref{CC}(a) and the
$\phi$-dependent "scalar" term connected with diagrams Fig.\ref{CC}(b,c):
\be V(\rLV,\phi,m^2,\la,\mu)=
V_{vac}(\rLV,m^2,\la,\mu) \label{Veff} + V_{scal}(\phi,m^2,\la,\mu) \,, \ee
where the parameter $\rLV=\rLV(\mu)$ depends on the vacuum cosmological constant and will be defined in the next section.
In order to understand the origin of this splitting, one introduces the functional called the effective action of the vacuum 
$\Ga_{vac}$. It is part of the full effective action which is left when
the mean scalar field $\phi$ is set to zero:
$\Ga_{vac}=\Ga[\phi=0]$. Thus, it is a pure quantum object
which only depends on the set of parameters $P=m,\lambda,...$
of the classical theory. At the functional level, the generating functional $W$ for the vacuum-to-vacuum transition amplitude is 
\be W[J=0]=e^{i\Ga_{vac}}\,=\,\int {\cal D}\phi \ e^{iS[\phi; J=0]}\,,
\label{vacEA2} \ee
where the source $J$ is set to zero. In this way, the functional
$\Ga_{vac}$ is the generator of the proper vacuum-to-vacuum
diagrams. 
%In flat space, these diagrams are removed by normalizing
%the functional, but in the curved space they are significant.

The RG-invariance of the
full renormalized effective potential reads (where $\gamma_m m^2 = \beta_{m^2}$):
\be 
\Big(\mu\frac{\pa}{\pa \mu}+\beta_{\la}\frac{\pa}{\pa \la} +\ga_m m^2
\frac{\pa}{\pa m^2} -\ga_\phi \phi \frac{\pa}{\pa \phi} +\beta_{\rLV}\frac{\pa}{\pa \rLV}\Big)V(\rLV,\phi,m^2,\la,\mu)\,=\,0\,. 
\label{RG0} 
\ee
%
%From the RG-invariance of the renormalized EA -- see
%Eq.\,(\ref{nn8}) -- it follows immediately the $\mu$-independence of
%the renormalized functional $\Ga_{vac}$ and, therefore, we arrive at
%the second identity (\ref{RG-sep}) for the vacuum part of the
%effective potential, while the first identity is the result of the
%subtraction of (\ref{RG-sep}) from (\ref{RG}). The net result is
%that the vacuum and matter parts of the effective potential are
%overall $\mu$-independent separately and no cancelation between them
%can be expected. \vskip 1mm
%
Using (\ref{Veff}), we now show that Eq.\,(\ref{RG0}) is, in fact, a sum of two independent RG equations,
\begin{eqnarray}
&&\Big(\mu\frac{\pa}{\pa \mu} + \beta_{\la} \frac{\pa}{\pa \la}+\ga_m
m^2 \frac{\pa}{\pa m^2}+\beta_{\rLV}\frac{\pa}{\pa \rLV}\Big)V_{vac}(\rLV,m^2,\la,\mu) \,=\,0\,,
\label{RG-sep0}
\\
&&\Big(\mu\frac{\pa}{\pa \mu} + \beta_{\la}\frac{\pa}{\pa \la}
+\ga_m m^2 \frac{\pa}{\pa m^2} -\ga_\phi \phi \frac{\pa}{\pa \phi}
\Big)V_{scal}(\phi,m^2,\la,\mu)\,=\,0\,.
\label{RG-scalar0}
\end{eqnarray}
To prove this, notice that from the RG-invariance of the renormalized effective action follows the $\mu$-independence of
the renormalized functional $\Ga_{vac}$ and, therefore, we arrive at
the first identity (\ref{RG-sep0}) for the vacuum part of the
effective potential, while the second identity is then the result of the
subtraction of (\ref{RG-sep0}) from (\ref{RG0}). We will illustrate this point later.

The net result is
that the vacuum and matter parts of the effective potential are
overall $\mu$-independent separately and no cancelation between them
is expected.

\subsection{Vacuum part of the CC}
Let us compute the $V_{vac}(\rLV,m^2,\la,\mu)$ object at the one-loop level. We start from
\begin{equation}
S_{HE}=-\frac{1}{16\pi G_{vac}^0}\int d^{4}x\sqrt{-g}\,\left\{ \,%
\,R+2\Lambda _{vac}^0 \,\right\} +S_{matter}\,.  \label{HE1}
\end{equation}
As it is well-known the $\Lambda _{vac}^0$-dependent part has exactly the form of the bare vacuum energy density (\ref{CCd3}) \footnote{Sometimes, in the literature ($\rho^0_{\Lambda})^{vac} = \frac{\Lambda^0_{vac}}{8\pi G^0_{vac}} \equiv h m^4$ is used where $h$ is treated as an independent parameter.}:
\be (\rho^0_{\Lambda})^{\rm vac} = \frac{\Lambda^0_{vac}}{8\pi G^0_{vac}} \,. \label{CCd2} \ee
In the standard QFT, the loop-divergent terms in the vacuum density are absorbed by the bare cosmological constant term $(\rho^0_{\Lambda})^{vac}$ of the Hilbert-Einstein action. For this, we split the bare term $(\rho^0_{\Lambda})^{vac} $
as 
\be (\rho^0_{\Lambda})^{\rm vac} =  \rho_{\Lambda}^{\rm vac}(\mu)+\delta\rLV \,,
\label{counter}
\ee 
where the counterterm $\delta\rLV$ depends on the regularization and
the renormalization scheme.
Specifically, the one-loop effects encoded in ${\bar V}_{vac}^{(1)}$ modify this relation as follows:
\begin{equation}
\label{Vacfree} V_{vac}(\rLV,m^2,\la,\mu)= (\rho^0_{\Lambda})^{vac}+{\bar V}_{vac}^{(1)} =\rLV(\mu)+\delta\rLV+{\bar V}_{vac}^{(1)}\,.
\end{equation}

Now, starting from (\ref{vacEA2}), the variation of vacuum-to-vacuum transition amplitude with mass $m$ leads to the (Euclidean) Green's function at coincident points \cite{BrownBook}
\be
\frac{\pa }{\pa m^2} \log W[J=0] =-\frac{1}{2} \int d^n_E x \ \Delta_E(0) \quad \text{with} \quad \Delta_E(0) =  \int \frac{d^n_E k }{(2\pi)^n}\frac{1}{k^2+m^2}=\frac{(m^2)^{\frac{n}{2}-1}}{(4\pi)^{n/2}}\Gamma(1-\frac{n}{2}) \ .\nonumber \\
\ee
In terms of Feynman diagrams this is just the vacuum bubble shown in Fig.\ref{CC}(a). Integrating this equation and adding $(\rho^0_{\Lambda})^{vac}$ we obtain the vacuum energy density (\ref{Vacfree}) as:
\be
W[J=0]= e^{-  \int d^n_E x  \ V_{vac}^{}} \quad \text{with} \quad  V_{vac}^{}=(\rho^0_{\Lambda})^{\rm vac}+{\bar V}_{vac}^{(1)} =(\rho^0_{\Lambda})^{\rm vac}
+\frac{m^n}{(4\pi)^{n/2}}\frac{1}{n}\Gamma(1-\frac{n}{2}) \label{Vvac} \,.
\ee
The pole of the Gamma function in 4 dimensions $\Gamma(1-\frac{n}{2})\sim 2/(n-4)$ so for $n\rightarrow 4$:
\begin{equation}\label{Vacfree2}
{\bar V}_{vac}^{(1)}
 %= \frac12\,\mu^{4-n}\, \int\frac{d^{n-1} k}{(2\pi)^{n-1}}
   %\,\sqrt{\vec{k}^2+m^2}
= \frac{m^4}{64\pi^2}\,\left(\frac{2}{n-4}
-\ln\frac{4\pi\mu^2}{m^2}+\gamma_E-\frac32\right) \,.
\end{equation}
Equation (\ref{Vacfree2}) is divergent and needs a subtraction. If
we adopt the $\overline{MS}$ subtraction scheme, the counterterm
$\delta\rLV$ gets fixed in such a way that the renormalized vacuum energy density at 1-loop
is
\begin{equation}
\label{rfvacenergy2}
V_{vac}(\rLV,m^2,\la,\mu)\,=\,\rLV(\mu)+ \delta\rLV+{\bar V}^{(1)}_{vac}
\,=\,\rLV(\mu)+\frac{m^4}{64\pi^2}\,
\,\left(\ln\frac{m^2}{\mu^2}-\frac32\right)\,.
\end{equation}
This is the result for $V_{vac}(m^2,\la,\rLV,\mu)$ at 1-loop. Notice that it is a pure quantum object that (to one-loop order) depends only on the parameter $m$ of the
classical Lagrangian and does not depend (to this order) on $\lambda$.

It is clear from (\ref{rfvacenergy2}) that the cosmological constant is renormalized according to:
\be
(\rho^0_{\Lambda})^{\rm vac} =\mu^{n-4} \bigg(\frac{m^4}{2(4\pi)^2}\frac{1}{n-4}+ \rho_{\Lambda}^{\rm vac}(\mu)\bigg) \qquad 
%\Longrightarrow \qquad
\ee
and 
\be
\mu \frac{\partial \rLV}{\partial \mu}\equiv \beta_{\rLV} =\frac{m^4}{32\pi^2} \ .
\label{MSrenom}
\ee 
This is the expression for $\beta_{\rLV}$ calculated in the $\overline{MS}$ scheme.

\begin{figure}[h]
 \center
\includegraphics[width=0.2\textwidth]{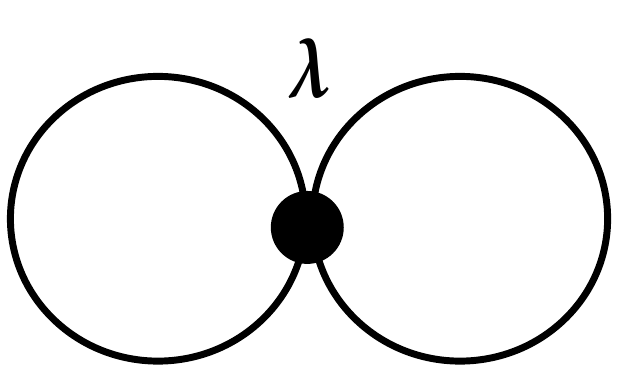}
\caption{The two-loop contribution to the cosmological constant in $\phi^4$ theory.
\label{CC2loop}}
\end{figure}

In writing this equation we used the renormalized mass because the RG equations must involve only finite (renormalized) quantities. However, so far, we only computed the vacuum bubble in the free theory, where renormalized and bare masses are the same. In the interacting theory, we have to take care about the renormalization of the mass $m$ itself. The leading mass correction is based on the correction to the scalar propagator shown in Fig.\ref{CC}(b) and after the standard calculation we arrive at:
\be
m_0^2 = m^2 \bigg( 1- \frac{3\lambda}{(4\pi)^2}\frac{1}{n-4} \bigg) \qquad \Longrightarrow \qquad 
\mu \frac{\partial m^2}{\partial \mu}=m^2 \frac{3\lambda}{(4\pi)^2} \ .
\label{masscorr}
\ee
The addition of the interactions modifies the renormalization of the cosmological constant according to the two-loop diagram shown in Fig.\ref{CC2loop} 
which leads to:
\be
V_{vac}^{}=(\rho^0_{\Lambda})^{vac}+\frac{m_0^n}{(4\pi)^{n/2}}\frac{1}{n}\Gamma(1-\frac{n}{2}) 
+\frac{3\lambda_0}{8}\Bigg[\frac{m_0^{n-2}}{(4\pi)^{n/2}}\Gamma(1-\frac{n}{2})\Bigg]^2 \,.
\ee
To put this expression in the renormalized form we have to replace the bare mass by the renormalized one using (\ref{masscorr}), while we can use renormalized quartic coupling $\lambda_0=\lambda$ to this order. One obtains:
\be
V_{vac}=(\rho^0_{\Lambda})^{vac}+\frac{m^4}{2(4\pi)^{n/2}}\frac{\mu^{n-4}}{n-4}\bigg( 1- \frac{3\lambda}{(4\pi)^2}\frac{1}{n-4} \bigg) +{\rm finite}
\label{newCC}
\ee
where we observe that the leading term is written in terms of the renormalized mass.
It is clear from (\ref{newCC}) that the cosmological constant is then renormalized according to:
\be
(\rho^0_{\Lambda})^{vac} =\mu^{n-4} \bigg[\frac{m^4}{2(4\pi)^2}\frac{1}{n-4}\bigg( 1- \frac{3\lambda}{(4\pi)^2}\frac{1}{n-4} \bigg)+ \rho_{\Lambda}^{vac}(\mu)\bigg] \; \Longrightarrow \quad 
 \beta_{\rLV} =\frac{m^4}{32\pi^2}\Bigg[1+\mathcal{O}\Big(\frac{\lambda}{16\pi^2}\Big)\Bigg]\ . \nonumber\\
\label{MSrenom1}
\ee 
Note again that there is no correction to the RG to the leading order in $\lambda$. Basically, each of the two bubbles in Fig.\ref{CC2loop} acts as a mass correction to the other one and gets reabsorbed into the renormalized mass.

%Plugging (\ref{MSrenom}) into (\ref{Vvac}) we obtain:
%\be
%
%\ee
%where, for our example with a real scalar field with mass $m$, the dimensionally regularized
%form of the vacuum-to-vacuum diagram in flat space gives
%where $n\rightarrow 4$ in the last expression. Equation (\ref{Vacfree2}) is divergent and needs a subtraction. If
%we adopt the $\overline{MS}$ subtraction scheme, the counterterm
%$\delta\rLV$ gets fixed in such a way that the renormalized vacuum energy density at 1-loop
%is
%\begin{equation}
%\label{rfvacenergy2}
%V_{vac}(m^2,\la,\rLV,\mu)\,=\,\rLV(\mu)+ \delta\rLV+{\bar V}^{(1)}_{vac}
%\,=\,\rLV(\mu)+\frac{m^4}{64\pi^2}\,
%\,\left(\ln\frac{m^2}{\mu^2}-\frac32\right)\,.
%\end{equation}
%This is the result for $V_{vac}(m^2,\la,\rLV,\mu)$ at 1-loop. Notice that it is a pure quantum object that (to one-loop order) depends only on the parameter $m$ of the
%classical Lagrangian and does not depend (to this order) on $\lambda$.

\subsection{Decoupling effects}

%Recall that 
%%we could simply apply (\ref{RG-sep0}) to (\ref{rfvacenergy2}) to obtain
%\be
%\mu \frac{\partial \rLV}{\partial \mu}\equiv \beta_{\rLV}=\frac{m^4}{32\pi^2}\ \ .
%\label{aaa}
%\ee
By definition, the RG equation (\ref{MSrenom1}) holds in the region $\mu\gg m$ and to go to the opposite regime $\mu\ll m$ would require to take into account:
1) the contribution of heavy particles at the energies near their
mass, 2) the ÒresidualÓ effects from the heavy particles at energies well below their mass.

It is well-known that the decoupling of heavy particles does not hold in a mass-independent scheme like the $\overline{MS}$,
and for this reason they must be decoupled by hand using the sharp cut-off procedure or some of the mass-dependent schemes.
The quantum effects of the massive particles are, in principle, suppressed at low energies by virtue of the Appelquist-Carazzone theorem \cite{Appelquist:1974tg}, so
that in the region below the mass of the particle its quantum effects become smaller. At this point we need the relation between the IR and the
UV regions which would require to extend the Wilson RG for the quantitative
description of the threshold effects, and to apply a mass-dependent RG formalism.
%But, since both of these formalisms are too cumbersome for an investigation at this stage, we will simply start with arguments of the dimensional analysis.

%first of all tackle the problem by applying the standard Òsharp cut-offÓ approximation within the
%minimal subtraction (MS) scheme, namely the contribution of a particle will be taken into account
%only at the energies greater than the mass of this particle

%The first  "vacuum bubble" diagram Fig.\ref{CC}a contributes to $\Lambda _{vac}(\mu)$ and the result for the $\beta$-function
%in the presence of arbitrary degrees of freedom of spin $J$ and
%non-vanishing mass $M_{J},$ reads
%\begin{equation}
%\beta _{\Lambda_{vac} }=
%\frac{1}{2}Str [M_J^4]
%=\,(-1)^{2J}(J+1/2)\,n_{c}\,n_{J}\,\,M_{J}^{4} \ , \qquad Str\equiv \sum\limits_{scalars}-\ 2\sum\limits_{Weyl fermions}+\ 3\sum\limits_{vectors} 
%\label{betalambda}
%\end{equation}
%$\,$with $\,(n_{c},n_{1/2})=(3,2)\,$ for quarks, $(1,2)$ for leptons and $%
%(n_{c},n_{0,1})=(1,1)$ for scalar and vector fields. 

%At lower energies
%$\mu \ll M_J^{heavy}$, heavy particles of some mass $M_J^{heavy}$ are in principle
%suppressed by virtue of the Appelquist-Carrazone theorem.
%However,  it would be premature to claim the validity of this decoupling theorem because
On purely dimensional grounds, in the regime $\mu\ll m$ one expects the corrections to the CC
of the type $\mu^2 m^2$.
%In this case the modifications of the previous
%picture will also depend on the possible choices for the RG scale $\mu $
%which, as we have seen in Section 2, is not a completely obvious matter in
%the cosmological\ scenario. Whether these heavy mass terms are eventually
%relevant or not is not known, but one can at least discuss this possibility
%on generic grounds. Under this hypothesis the heavy particle effects
%emerging in\ a mass-dependent RG scheme would lead to a new RG equation for $%
%\Lambda _{ph}$ in which the \textit{r.h.s.} of eq. (\ref{RGEMF}) ought to be
%modified by additional dimension-$4$ terms involving the scale $\mu $
%itself, namely terms like $\mu ^{2}M^{2}$. \ 
These corrections can be seen from the
fact that in a mass-dependent subtraction scheme a
heavy mass $m$ enters the $\beta $-functions through the dimensionless
combination $\mu /m$, so that the CC, being a dimension-$4$ quantity, is
expected to have the $\beta $-function corrected as follows:
\begin{equation}
\beta \big(m_{light},\frac{\mu }{m}\big)=a\,m_{light}^{4}+b\,\left( \frac{\mu }{m}\right) ^{2}m^{4}+ 
c \,\left( \frac{\mu }{m}\right) ^{4}m^{4}%
+\mathcal{...} 
\label{betaM}
\end{equation}
where $a$,$b$ and $c$ are some coefficients, $m_{light}$ is some light mass $m_{light} \ll \mu$, and the dots stand for terms suppressed
by higher order powers of $\mu /m \ll 1$.  Equipped with the necessary RG formalism and expectation of decoupling behavior based on the dimensional analysis, 
we will show how one can deal with the decoupling effect in the full SM and how to calculate explicitly the coefficients $a,b,c$ {\it for any model}.

\section{RG running of the Cosmological Constant in the Standard Model} \label{SMRG}

Before discussing the mass-dependent RG schemes relevant for decoupling, let us recall the results in the usual $\overline{MS}$ scheme. 
\subsection{Mass-independent ($\overline{MS}$) scheme}

%\subsection{Mass-independent ($\overline{MS}$) scheme in flat space}

The renormalized effective potential of the SM, $V$, can
be written in the 't Hooft-Landau gauge and the $\overline{MS}$ scheme as \cite{Casas:1994qy,Ford:1992mv}
\be
\label{veff}
V(\rLV,\phi, m^2,\lambda_i,\mu)\equiv V_0 + V_1 + \cdots \;\; ,
\ee
where $\lambda_i\equiv (g,g',\lambda,h_t)$ runs over all
dimensionless couplings and $V_0$, $V_1$ are 
the tree level potential and the one-loop correction respectively, namely
%\subequations{
\be
\label{v0}
V_0=-{\displaystyle\frac{1}{2}}m^2\phi^2 +
{\displaystyle\frac{1}{8}}\lambda\phi^4,
\ee
\be
\label{v1}
V_1={\displaystyle\sum_{i=1}^5}{\displaystyle\frac{n_i}{64\pi^2}}
M_i^4(\phi)\left[\log{\displaystyle\frac{M_i^2(\phi)}{\mu^2}}
-c_i\right]+\rLV \ ,
\ee
%}
%\endsubequations
with \footnote{There is a logarithmic singularity associated with massless particles. In the SM, it is well-known that in the Landau gauge the Goldstone boson mass,
 which vanishes at the minimum of the effective potential, presents an infrared logarithmic
divergence for the running Higgs mass. However, the physical mass $M_{phys}$ (corrected by the self-energy shift from $p^2=0$ to $p^2=M_{phys}^2$) is finite since the divergent contribution to the running mass coming from the Goldstone bosons is cancelled by the contribution of the Goldstones to the self-energy \cite{Casas:1994us}.}
\be
M_i^2(\phi)=\kappa_i\phi^2-\kappa_i^m m^2,
\ee
and coefficients $n_i$, $\kappa_i$, $\kappa_i^m$, and $c_i$ defined in Table.\ref{SMmassesFlat}.
\begin{table}
\center
\caption{\label{SMmassesFlat}Contributions to the effective potential (\ref{v1})
 from the SM particles $W^{\pm}$, $Z^0$, top quark t, Higgs $\phi$ and the Goldstone bosons $\chi_{1,2,3}$.}
%\vspace{2mm}
 \begin{tabular}{|c||ccccc|}
 \hline
   $\Phi$ & $~~i$ & $~~n_i$  &$~~\kappa_i$ & $\kappa_i^m$            & $\quad c_i~~$ \\[1mm]\hline
   $~W^\pm$ & $~~1$  & $~~6$       & $~~ g^2/4$        & $0$          & $\quad{5}/{6}~~$ \\[1mm]
  $Z^0$ & $~~2$  & $~~3$       & $~~(g^2+g'^2)/4$ & $0$       & $\quad{5}/{6}~~$ \\[1mm] \hline
  t & $~~3$  & $-12$     & $~~ y_{\rm t}^2/2$      & $0$         & $\quad{3}/{2}~~$ \\[1mm]\hline
   $\phi$ & $~~4$  & $~~1$       & $~~3\lambda/2$           & $~1$     & $\quad{3}/{2}~~$ \\[1mm]\hline
   $\chi_i$ & $~~5$  & $~~3$       & $~~\lambda/2$            & $~1$    &   $\quad{3}/{2}~~$\\[.5mm]\hline
  \end{tabular}
  \end{table}
%\bear{llll}
%n_1=6\ ,&  \kappa_1={\displaystyle\frac{1}{4}}g^2\ ,& \kappa_1^m=0 \ ,&
%c_1={\displaystyle\frac{5}{6}}\ ;\vspace{0.3cm}\\
%n_2=3\ ,&  \kappa_2={\displaystyle\frac{1}{4}}[g^2+g'^2]\ ,&
%\kappa_2^m=0\ ,& c_2={\displaystyle\frac{5}{6}}\ ;\vspace{0.3cm}\\
%n_3=-12\ ,& \kappa_3={\displaystyle\frac{1}{2}}h_t^2\ ,& \kappa_3^m=0\ ,&
%c_3={\displaystyle\frac{3}{2}}\ ;\vspace{0.3cm}\\
%n_4=1\ ,&  \kappa_4={\displaystyle\frac{3}{2}}\lambda\ ,
%& \kappa_4^m=m^2\ ,&
%c_4={\displaystyle\frac{3}{2}}\ ;\vspace{0.3cm}\\
%n_5=3\ ,&  \kappa_5={\displaystyle\frac{1}{2}}\lambda\ ,
%& \kappa_5^m=m^2\ ,&
%c_5={\displaystyle\frac{3}{2}}\ .\vspace{0.3cm}
%\label{SMmasses}
%\eear
$M_i^2(\phi)$ are the tree-level expressions for the background-dependent masses of the
particles that enter in the one-loop radiative corrections, namely
$M_1\equiv m_W$, $M_2\equiv m_Z$, $M_3\equiv m_t$, $M_4\equiv
m_{\rm Higgs}$, $M_5\equiv m_{\rm Goldstone}$.
The parameter $\rLV(\mu)$ is the SM analogue of the renormalized cosmological constant $\rLV(\mu)$ for the real scalar field discussed in the previous section. Repeating the procedure as before, 
%As discussed in the previous section 
we split the effective potential into two pieces: the $\phi$-independent (vacuum) term and the
$\phi$-dependent "scalar" term
\be
V(\rLV,\phi,m^2,\la_i,\mu)=V_{vac}(\rLV,m^2,\la_i,\mu)  + 
V_{scal}(\phi,m^2,\la_i,\mu)
 \ .
\ee
Various pieces satisfy the RG equations (\ref{RG0}) and (\ref{RG-scalar0},\ref{RG-sep0}) with $\lambda\to \lambda_i$ and these equations are valid for any value of $\phi$.

However, for the extremum value $\phi=\langle\phi \rangle$ defined via 
 $\frac{\partial V_{scal}(\phi)}{\partial \phi}\Big |_{\langle\phi \rangle}=0$,  the term containing anomalous dimension of the Higgs $\gamma_\phi$ drops out and (\ref{RG-scalar0}) reads:
\begin{eqnarray}
&&\Big(\mu\frac{\pa}{\pa \mu} + \beta_{\la_i}\frac{\pa}{\pa \la_i}
+\ga_m m^2 \frac{\pa}{\pa m^2} 
\Big)V_{scal}(\langle\phi \rangle,m^2,\la,\mu)\,=\,0\, .
\label{RG-scalar1}
\end{eqnarray}
Using the tree-level potential (\ref{v0}), it is useful to define parameter 
\be\rho_{ind}(\mu)\equiv V_0 (\langle\phi \rangle)=-\frac{m^4(\mu)}{2\lambda(\mu)} \ .
\ee
%playing the role similar to the $\rLV(\mu)$. 
The running of this parameter reads:
\be
\beta_{\rho_{ind}}\equiv \mu\frac{\pa \rho_{ind}(\mu)}{\pa \mu}=\rho_{ind}(\mu) \  \Big(2\gamma_m -\frac{\beta_{\la}}{\la} \Big) \ .
\label{ind1}
\ee

 %$\Lambda_{ind}\equiv V_{scal}(\langle\phi \rangle,m^2,\la_i,\rLV,\mu)$

%\cite{Einhorn},
%which will turn to be irrelevant in our calculation.
By equating the terms with the different powers of $\phi$ : 
\be
\mu\frac{dV}{d \mu}\sim \big( \ \phi^4 [...]+ m^2 \phi^2[....]+ m^4[...]\ \big)=0 \ ,
\ee
 it is straightforward to check that the requirement (\ref{RG0})
applied to the full one-loop effective potential (\ref{veff})
leads to 
%\subequations{
\be
{\displaystyle\frac{1}{8}}\beta_{\lambda}-{\displaystyle\frac{1}{2}}
\gamma_\phi\lambda=
{\displaystyle\sum_i}{\displaystyle\frac{n_i\kappa_i^2}{32\pi^2}} \ ,
\label{p4inv}
%\vspace{0.3cm}
\ee
\be
{\displaystyle\frac{1}{2}}\gamma_{m}-\gamma_\phi=
{\displaystyle\sum_i}{\displaystyle\frac{n_i\kappa_i\kappa_i^{m}}{16\pi^2}} \ ,
\label{p2inv}
%\vspace{0.3cm}
\ee
\be
{\displaystyle \mu \frac{\partial \rLV}{\partial \mu}}=m^4{\displaystyle\sum_i}{\displaystyle\frac{n_i(\kappa_i^{m})^2}{32\pi^2}}
\label{p0inv} \ ,
\ee
%}
%\endsubequations
up to two-loop corrections. The first two equations come from $\phi^4 [...]$ and $m^2 \phi^2[....]$ terms respectively and they belong to (\ref{RG-scalar0}) and the last condition came from $m^4[...]$ and satisfy (\ref{RG-sep0}). These equations show explicitly that the vacuum $V_{vac}$ and scalar $V_{scal}$ parts of the
full effective potential satisfy independent RG equations. 

%For the extremum value $\phi=\langle\phi \rangle$ we have to drop the $\gamma_\phi$ terms from  and subtracting these equations appropriately we obtain the running of the $\rho_{ind}$ :
Subtracting (\ref{p4inv}) from (\ref{p2inv}) appropriately, we reconstruct \ref{ind1}
\be
\beta_{\rho_{ind}}=\rho_{ind}\Bigg({\displaystyle 2\gamma_{m}}-{\displaystyle\frac{\beta_{\lambda}}{\la}}\Bigg)=
m^4\Bigg({\displaystyle\sum_i}{\displaystyle\frac{n_i\kappa_i^2}{8\pi^2\la^2}} - {\displaystyle\sum_i}{\displaystyle\frac{n_i\kappa_i\kappa_i^{m}}{8\pi^2\lambda}} \Bigg ) \ .
\label{ind}
\ee
Combining (\ref{ind}) and (\ref{p0inv}) we finally obtain 
%(remember, $\kappa_4^m=\kappa_5^m=m^2$ ):
\be
\boxed{{\displaystyle \mu \frac{\partial (\rLV + \rho_{ind})}{\partial \mu}}=m^4\Bigg({\displaystyle\sum_i}{\displaystyle\frac{n_i\kappa_i^2}{8\pi^2\la^2}} - {\displaystyle\sum_i}{\displaystyle\frac{n_i\kappa_i\kappa_i^{m}}{8\pi^2\lambda}} +{\displaystyle\sum_i}{\displaystyle\frac{n_i(\kappa_i^{m})^2}{32\pi^2}}\Bigg )={\displaystyle\sum_{i}^{}}{\displaystyle\frac{n_i}{32\pi^2}} M_i^4(\langle\phi \rangle)} \ ,
\label{central}
\ee
where we used 
\be
M_i^2(\langle\phi \rangle)=\kappa_i\langle\phi \rangle^2-\kappa_i^m m^2 = m^2 \left (\kappa_i \frac{2 }{\lambda} - \kappa_i^m \right )  
\label{massextr}
\ee
and $\langle\phi \rangle^2 = 2 m^2/\lambda$. 
Eq.(\ref{central}) is the central equation valid \emph{in the UV regime} of massless and massive theories, theories with the spontaneous symmetry breaking (SSB) and without. This equation defines, in a compact form, the total running of the implicit $\mu-$dependences on the l.h.s. by balancing them with the explicit $\mu-$dependences on the r.h.s \cite{Shapiro:2008yu}.

%Going from the $\overline{MS}$ masses to the physical (pole) masses we obtain:
%\be
%M_t^4(\langle\phi \rangle)=(M^{ph}_t)^4\bigg(1-4 C_F\frac{\alpha_s}{\pi}+\frac{8 y_t^2}{(4 \pi)^2}\bigg) \qquad M_h^4(\langle\phi \rangle)=(M^{ph}_h)^4 \bigg(1+\frac{4 y_t^2}{(4 \pi)^2}\bigg)
%\label{pole}
%\ee
%where $C_F=4/3$ and $\overline{MS}$ coupling constants $\alpha_s$ and $y_t$ are evaluated at the physical top mass.

\subsection{Mass-dependent scheme} \label{MD1}

Now, following 
%\cite{Babic:2001vv} and
the discussion above, we may generalize approach to the mass-dependent RG scheme. As we discussed above the decoupling of heavy particles does not hold in a mass-independent $\overline{MS}$ scheme and here we recall how to get around this problem. 

\begin{figure}[h]
\includegraphics[width=0.95\textwidth]{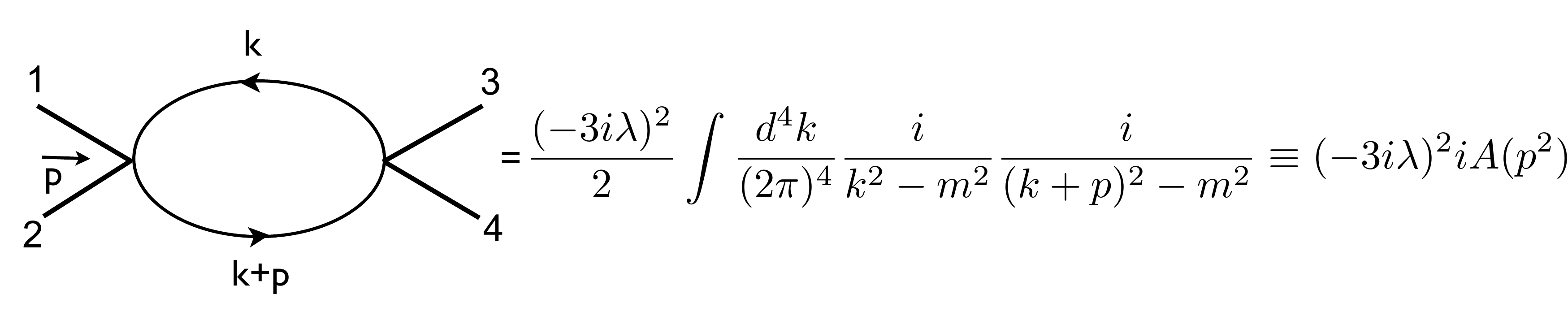}
\caption{\em
(\textit{a}) The one-loop contributions to 4-point function in $\phi^4$ theory.
\label{Pole4}}
\end{figure}

The basic issue can be seen in the computations of 2$\to$ 2 scattering amplitude in a simple $\phi^4$-theory:
\be
V_{\phi^4}=-{\displaystyle\frac{1}{2}}m^2\phi^2 +
{\displaystyle\frac{1}{8}}\lambda\phi^4
\ee
which is just the potential of (\ref{v0}) limited to one real scalar. The exemplary scattering amplitude is shown in Fig.\ref{Pole4}, where $p=p_1+p_2$ is the total incoming momenta. Expanding in terms of the external momentum $p$, it is only the term $p=0$ which is
divergent since every power of $p$ effectively gives one less power of $k$
for large $k$.

Computing the logarithmically-divergent integral using, for example, dimensional regularization, we obtain:
\be
A(p^2)= -\frac{1}{32 \pi^2}\int_0^1 dx \ \Bigg(\frac{2}{\epsilon} -\gamma_E+\log(4\pi) -\log[m^2-x(1-x)p^2]\Bigg)
\label{ampl}
\ee
where we see explicitly that $p^2$-terms are finite. 

In the $\overline{MS}$ renormalization scheme, one chooses counterterms ({\it c.t.}) in such a way as to remove the divergent $2/\epsilon$ pole and scale independent number $-\gamma_E+\log(4\pi) $ and therefore, by construction, the counterterms are \emph{mass-independent}. Also one introduces the arbitrary mass parameter $\mu_{\overline{MS}}$ to make equation dimensionally correct so that finally:  
%\footnote{Formally,  we can think of making an additional finite subtraction at some arbitrary energy (squared) scale $-\mu^2_{\overline{MS}}$ when choosing the $c.t.$ . See also the $\log(m^2/\mu^2)$ term in (\ref{Vacfree2}).  } 
 \be
A(p^2)_{\overline{MS}}=A(p^2)+c.t.= \frac{1}{32 \pi^2}\int_0^1 dx \ \log\Bigg(\frac{m^2-x(1-x)p^2}{\mu_{\overline{MS}}^2}\Bigg) \,.
\label{version1}
\ee
From the RG equation applied to the 4-point function:
\be
G_{\overline{MS}}^{(4)}(m^2,\la,\mu_{\overline{MS}})\sim -3i\lambda+9(-i\lambda)^2 [i A(s)_{\overline{MS}}+i A(t)_{\overline{MS}}+i A(u)_{\overline{MS}}]
\ee 
one now derives the $\beta$-function of the theory:
\be
\Big(\mu_{\overline{MS}}\frac{\pa}{\pa \mu_{\overline{MS}}} + \beta_{\la}^{\overline{MS}}\frac{\pa}{\pa \la}
\Big)G^{(4)}(m^2,\la,\mu_{\overline{MS}})\,=\,0\,  \qquad \Longrightarrow \qquad \beta_{\lambda}^{\overline{MS}}= \frac{9\lambda^2}{16\pi^2} \ .
\ee

In a mass-dependent renormalization scheme, the counterterms are \emph{mass-dependent} and can be chosen, for example, to subtract from (\ref{ampl}), in addition to the divergent pole and scale-independent numbers, also the $\log[m^2-x(1-x)p^2]$ evaluated at the $p^2=-\mu^2$ where $\mu$ is yet another arbitrary scale. After this additional \emph{finite subtraction}, (\ref{ampl}) will be replaced by the corresponding expression in the momentum subtraction scheme (MOM) as 
\be
A(p^2)_{\MS}=A(p^2)+c.t.= \frac{1}{32 \pi^2}\int_0^1 dx \ \log\Bigg(\frac{m^2-x(1-x)p^2}{m^2+x(1-x)\mu^2}\Bigg) \ .
\label{version2}
\ee
%and we observe the correspondence between the arbitrary scales as: $\mu_{\overline{MS}}^2=m^2+x(1-x)\mu^2$. 
Again, $\mu-$dependence will determine the beta function of the theory through the RG-equation and we obtain:
\be
\Big(\mu\frac{\pa}{\pa \mu} + \beta^{\MS}_{\la}\frac{\pa}{\pa \la}
\Big)G_{\MS}^{(4)}(m^2,\la,\mu)\,=\,0\,  \quad \Longrightarrow \quad \beta_{\lambda}^{\MS}= \frac{9\lambda^2}{16\pi^2}\int_0^1 \frac{x(1-x)\mu^2}{m^2+x(1-x)\mu^2} dx \nonumber \\
\label{betaMD}
\ee
which in the $\mu\ll m$ region, reproduces the decoupling behavior $\mu^2/m^2$ we discussed in (\ref{betaM}).

%Similarly, now we need to find the appropriate scheme transformation from the $\overline{MS}$ effective potential (\ref{veff1}) to the mass-dependent scheme which incorporates the decoupling effects. First, we notice that in the effective potential the role of the parameter $m$ in the scalar propagator in the loop in Fig.\ref{Pole4} is played by the background-dependent masses $M_i(\phi)$ of the SM particles defined in (\ref{Rmasses}). Second, 
Now, we need to generalize the above derivation in the mass-dependent scheme for the simple $\phi^4$-theory to the full SM including the loops of the $W,Z,t$ and Goldstones. 
%in the 2$\to$ 2 scattering in Fig.\ref{Pole4}.
 In Appendix A, we show that the appropriate generalization of (\ref{p4inv}) is given by:
\be
\left[{\displaystyle\frac{\beta_{\lambda}}{8}}-{\displaystyle\frac{\gamma_\phi\lambda}{2}}\right]_{\MS}=
{\displaystyle\sum_i}{\displaystyle\frac{n_i\kappa_i^2}{32\pi^2}}\int_0^1 \frac{x(1-x)\mu^2 dx}{(M_{phys }^2)_i+x(1-x)\mu^2}
\vspace{0.3cm}
\label{scheme}
\ee
in agreement with \cite{Spencer-Smith:2014woa}. For the single real scalar case discussed above, we have to 1-loop $\gamma_\phi=0$ and, from Table \ref{SMmassesFlat} we have $n_i=1$ and $\kappa_i=3\lambda/2$ so that we reproduce (\ref{betaMD}). Notice that when performing the sum over Goldstones, the parameter $(M_{phys }^2)_i$ becomes the physical mass of the vector boson corresponding to the Goldstone of type $i$.

Similarly, in Appendix A we also show that the generalization of (\ref{p2inv}) is given by: 
%-(\ref{p2inv3}) where now we have a corresponding equations with the r.h.s. multiplied by the same integral as in (\ref{scheme}).
\be
\left[{\displaystyle\frac{1}{2}}\gamma_{m}-\gamma_\phi\right]_{\MS}=
{\displaystyle\sum_i}{\displaystyle\frac{n_i\kappa_i\kappa_i^{m}}{16\pi^2}} \int_0^1 \frac{x(1-x)\mu^2 dx}{(M_{phys }^2)_i+x(1-x)\mu^2} \ .
\label{schemem}
\ee
We therefore conclude, that in this mass-dependent scheme, the corresponding MOM expression for the $\overline{MS}$ running of $\rho_{ind}$ in (\ref{ind}) takes the following form:
\be
\mu\frac{\pa \rho_{ind}(\mu)}{\pa \mu}_{|\rm MOM}=\rho_{ind}(\mu) \  \Big(2\gamma_m -\frac{\beta_{\la}}{\la} \Big) =\frac{ \langle\phi \rangle^4 }{32\pi^2}\sum_{i} n_i \big[\kappa_i^2- \kappa_i \kappa_i^m \lambda\big]\int_0^1 \frac{x(1-x)\mu^2 dx}{(M_{phys }^2)_i+x(1-x)\mu^2}.\nonumber \\ 
\label{indrun}
\ee

Now it remains to derive the vacuum part, Eqs.(\ref{MSrenom}) and (\ref{p0inv}), in the mass-dependent scheme. To accomplish that, one starts from the simple observation that the expression for the unrenormalized vacuum density (\ref{Vacfree2}) can be brought to the following form: 
\begin{equation}\label{Vacfree2A0}
{\bar V}_{vac}^{(1)}
= - \frac{m^4}{64\pi^2}\,\left( \frac{A_0(m)}{m^2} + \frac{1}{2}\right) \,.
\end{equation}
%
%\begin{equation}
%\label{rfvacenergy2}
%V_{vac}(m^2,\la,\rLV,\mu)\,=\,\rLV(\mu)+ \delta\rLV+{\bar V}^{(1)}_{vac}
%\,=\,\rLV(\mu)+\frac{m^4}{64\pi^2}\,
%\,\left(\ln\frac{m^2}{\mu^2}-\frac32\right)\,.
%\end{equation}
In above, $A_0(m)$ is the one-point Passarino-Veltman function with the properties given in the Appendix A.  
Now, using the relation 
\be
\mu\frac{\pa A_0 (M)}{\pa \mu}_{|\rm MOM}= M^2 \mu\frac{\pa B_0 (0,M,M)}{\pa \mu}_{|\rm MOM}=M^2 \mu\frac{\pa B_0 (p,M,M)}{\pa \mu}_{|\rm MOM}  = 2 M^2 \int_0^1 \frac{x(1-x)\mu^2 dx}{M^2+x(1-x)\mu^2}\nonumber \\ 
\label{A0B0} 
\ee
and the fact that $V_{vac}(m^2,\la,\rLV,\mu)$ satisfies the RG equation (\ref{RG-sep0}) 
%in (\ref{rfvacenergy2}) is the $\mu$-independent quantity, 
we obtain for the running of the vacuum part (\ref{p0inv}) in the MOM scheme \footnote{The "vacuum bubble" in Fig.\ref{CC}(a) is independent of the external momentum.  In order to have an external momentum probe
one needs to consider this "vacuum bubble" with external fields, such as for example the graviton legs. Then, to obtain the beta function for the $\rLV$ in the MOM scheme, one has to repeat the same steps as in the $\phi^4$ theory above. 
First, one has to calculate the renormalization 
of the quantum corrections to the n-point function of gravitons, then make a finite subtraction of the value of this quantity at $p^2=-\mu^2$ and, finally,
calculate the derivative $\mu\pa/\pa \mu$ of the form-factors.  

This program was carried out in \cite{Gorbar:2002pw} for the contributions of the loop
of massive scalar to the propagator (2-point function) of the gravitational perturbation 
$h_{\mu\nu}$ on the flat background $g_{\mu\nu}=\eta_{\mu\nu}+h_{\mu\nu}$ with the result that in this approach one cannot  
reveal the beta functions for $\rLV$ and the form of the decoupling remained unclear.}
\begin{align}
{\displaystyle \mu \frac{\partial \rLV}{\partial \mu}}_{|{\rm MOM}}=m^4 {\displaystyle\sum_i}{\displaystyle\frac{n_i(\kappa_i^{m})^2}{32\pi^2}}\int_0^1 \frac{x(1-x)\mu^2 dx}{(M_{phys }^2)_i+x(1-x)\mu^2}
\label{p0inv111} \ .
\end{align}
The essence of (\ref{A0B0}) is to ensure that once the finite subtraction (defining the MOM scheme) was made for $\beta_{\rho_{ind}}$, \emph{the same} finite subtraction is made for $\beta_{\rLV}$.

Putting everything together, and using again $\langle\phi \rangle^2 = 2 m^2/\lambda$, we achieve the generalization of (\ref{central}) 
\be
\boxed{{\displaystyle \mu \frac{\partial (\rLV +\rho_{ind})}{\partial \mu}}_{|\rm MOM}={\displaystyle\sum_{i}^{}}{\displaystyle\frac{n_i}{32\pi^2}} (M_{phys }^4)_i \int_0^1 \frac{x(1-x)\mu^2 dx}{(M_{phys }^2)_i+x(1-x)\mu^2}} \ ,
\label{centralMOM}
\ee
to the mass-dependent scheme. The Eq.(\ref{centralMOM}) is {\it the master equation} describing the running of CC in any regime, non-decoupling and decoupling one,  
%\be
%{\displaystyle \mu \frac{\partial (\rho_{ind}+\rLV+\kappa R)}{\partial \mu}}={\displaystyle\sum_{i}^{}}{\displaystyle\frac{n_i  M_i^4(\langle\phi \rangle) }{32\pi^2}} \Bigg(1-\frac{2 M_i^2(\langle\phi \rangle)  }{\mu\sqrt{4 M_i^2(\langle\phi \rangle) +\mu^2}}\log \frac{\sqrt{4 M_i^2(\langle\phi \rangle) +\mu^2}+\mu}{\sqrt{4 M_i^2(\langle\phi \rangle) +\mu^2}-\mu}  \Bigg)
%%{\mu^2+F(m, \langle\phi \rangle, \lambda,R)}\ .
%\label{central11}
%\ee
which is valid \emph{both in the UV and the IR regime}. In the UV regime we recover (\ref{central}), while in the IR regime we obtain the running of $\rLV + \rho_{ind}$ including the decoupling effects associated with the mass thresholds as 
\be
\int_0^1 {\displaystyle\frac{x(1-x)\mu^2 dx}{(M_{phys }^2)_i+x(1-x)\mu^2}}= 
 \left\{\begin{array}{lll}
\quad \qquad \qquad 1 \qquad \qquad \qquad  \mu^2\gg (M_{phys }^2)_i\\
{\displaystyle \frac{\mu^2}{6(M_{phys }^2)_i}-\frac{\mu^4}{30 (M_{phys }^4)_i}  \qquad \mu^2\ll (M_{phys }^2)_i}\\  
\end{array}\right. \,,
\ee
which will be used in the following to show the consequences of such running. 

%first equation provides the running of the parameter $\rho_{ind}$ including the decoupling effects associated with the mass thresholds while the second equation, due to the decoupling behavior in the running mass ( from (\ref{schemem}) we have $\gamma_m \sim \mu^2/m^2$ or $\beta_{m^2}\sim \mu^2)$ leads to the quick "freezeout" of $m=m(\mu)$ at its electroweak scale value so that $\beta_{\rLV}$ remains constant up to $\mu^2/m^2$ effects.%In principle, one can study the RG evolution of the $\rho_{ind}$, $\rLV$ and $\kappa$ parameters separately as they satisfy an independent RG equations and we will illustrate this separate RG running later. 
%However, in a generic mass-dependent RG scheme defined by the function $F(m, \phi, \lambda,R)$ the answer for various pieces might look quite messy since the r.h.s. of (\ref{central11}) will not be so simple anymore. Also, this form, is independent of how one reshuffles the constant pieces in the function $F(m, \phi, \lambda,R)$ between various CC contributions.

%\subsection{Final result} \label{Appl}
%We now generalize (\ref{centralMOM})
%give some examples for the application of the derived behavior of the CC (\ref{centralMOM}).
%\subsection{Standard Model}
%\begin{itemize}
%\item 
%{\bf Massive case $m\neq 0$:} 
In (\ref{centralMOM}) we derived the running of $\rLV+\rho_{ind}$ valid in the regime of decoupling of heavy particles $\mu^2\ll M_i^2(\langle\phi \rangle)$ while for the light particles $(m_{light}^2)_j \ll \mu^2$, we can simply use  (\ref{central}). Working in the region where $(m_{light}^2)_j \ll \mu^2\ll M_i^2(\langle\phi \rangle)$, we may combine these asymptotic results to obtain: 
\begin{eqnarray}
\label{central2}
&&\mu\frac{\pa (\rLV+\rho_{ind})}{\pa \mu}={\displaystyle\sum_{j}^{}}{\displaystyle\frac{n_j  (m_{light}^4)_j}{32\pi^2}}+
\frac{1}{32\pi^2}{\displaystyle\sum_i}n_i %\big[ m^4 (\kappa_i^{m})^2  
%-\lambda \langle\phi \rangle^4  \kappa_i(\kappa_i^m-\frac{ \kappa_i}{\lambda})\big]
(M_{phys }^4)_i
%-\frac{\lambda \langle\phi \rangle^4 }{32\pi^2}\sum_{i} n_i \kappa_i\ \big[\kappa_i^m-\frac{ \kappa_i}%{\lambda}\big]
\int_0^1 \frac{x(1-x)\mu^2 dx}{(M_{phys }^2)_i+x(1-x)\mu^2}\nonumber \\
=&&{\displaystyle\sum_{j}^{}}{\displaystyle\frac{n_j  (m_{light}^4)_j }{32\pi^2}}
%+\frac{m^4(\mu)}{8\pi^2}
+\frac{\mu^2}{12 (4\pi)^2}\Bigg[-12 M_t^2+6 M_{W}^2+3 M_Z^2+M_H^2
%12 m_t^2- 6 m_{W}^2-3 m_Z^2+\frac{m_H^4}{2}\Bigg(\frac{1}{m_W^2}+\frac{1}{2 m_Z^2}-\frac{3}{2m_H^2}\Bigg)
\Bigg] + \frac{\mu^4}{30 (4\pi)^2}
+\mathcal{O}\ \Bigg(\frac{\mu^6}{(M_{phys}^2)_i}\Bigg) \nonumber\\
\end{eqnarray}
The above expression is exactly of the form of (\ref{betaM}) and proves the expected decoupling behavior in the effective theories.
The light masses $m_{light}$ may be, again, generated by the Higgs vev $m_{light} (\langle\phi \rangle)$ such as a mass for, say, charm quark, or may be a new mass parameters in the SM Lagrangian related, for example, to the neutrino masses. As the $\mu$-scale slides down the energy, more and more SM masses will migrate from the $m_{light}^4$-term to the inside of the brackets in the $\mu^2$-term.

Notice that, when performing the sum over Goldstones, both beta functions for $\rLV$ (\ref{p0inv111}) and $\rho_{ind}$ (\ref{indrun}) have the term $\sim M_H^4(2/M_W^2+1/M_Z^2)$ but since it comes with the opposite sign it cancels in (\ref{central2}). This demonstrates the importance of considering the RG running of the total $\rLV+\rho_{ind}$ parameter rather than RG running of these contributions separately\footnote{ The $\mu^2$-term, up to overall numerical factor 1/6, is the result predicted in \cite{Babic:2001vv} for the beta function of $\rLV$ alone. 
We also disagree in the $\mu^4$-term 
which shows inconsistency of the derivation presented in \cite{Babic:2001vv}. 
%In contrast, this term was not present in the $\beta_{\rLV}$ presented in \cite{Babic:2001vv}.
}.

The $\mu^2 M_i^2$ term in the running of $\rLV + \rho_{ind}$ provides the leading RG effect due to the heavy SM particles \footnote{We are tacitly assuming that there is no $\mu^2 M^2$ contribution to the RG running from the particles decoupled at the higher scales such as Grand Unification or Planck scale.}  and we may demand it to vanish as to reduce the fine-tuning in the physical value of the CC at the $\mu _{c}=\mathcal{O}(10^{-3})\,$ eV. This requirement, however, leads to the SM prediction $m_H \approx 550$ GeV, inconsistent within the experimental value of $m_H\approx 125$ GeV.
%$m_H\approx 236$ GeV which is in contradiction with the experimental value of $m_H\approx 126$ GeV.

As discussed in \cite{Shapiro:2000dz}, heavy mass terms $\mu^2 M_i^2$
may also affect nucleosynthesis if we choose $\mu\sim T$, because they would induce vacuum energy density $\sim (T^2 M_i^2)/(4\pi)^2$ much bigger than the energy density of radiation $\rho_{rad}$ at the typical energy of nucleosynthesis $T\sim 10^{-4}$ GeV. On the other hand, the $m_{light}^4$- and $\mu^4$ - terms obey the constraint $\rho_{rad}<\rLV+\rho_{ind}$ in the energy interval relevant for nucleosynthesis. To avoid the problem, either we have to use alternative choice $\mu\sim H$ or, again, sufficient amount of fine-tuning should be arranged among the various $\mu^2 M_i^2$ terms. Since, as we saw, the heavy SM spectrum does not have this tuning, our results imply that SM has to be extended or $\mu\sim H$ choice is preferred over the $\mu\sim T$ \cite{Shapiro:2000dz}.

%For consistency with the nucleosynthesis, a next-to-leading $\mu^4$-term in (\ref{central2})
%has to be small too, \cite{Shapiro:2000dz} ....

%\item 
\section{Massless theories}\label{Appl}
We now apply our result (\ref{centralMOM}) to the massless theories which, as we will see, will give us new insights.

\subsection{Massless Standard Model\label{apl}}
In the massless limit of (\ref{central2}) $m=0$ ( {\it i.e.} all the terms with $\kappa_i^m$ absent), from (\ref{p0inv111}) we have $\rLV=$ const and only $\rho_{ind}$ runs with $\mu$. 
%Therefore, \emph{in the massless theories, we bypass the uncertainties related to the $\beta_{\rLV}$ in the MOM scheme.}
In this case, the $\mu^2$-term can be related to the Veltman condition as we now show. 

In the massless theory at the tree-level $\langle\phi \rangle=0$, which means that the tree-level mass of the Higgs is zero and the electroweak symmetry needs to be broken radiatively. For this to happen, we need to balance the tree-level potential against the 1-loop contribution, so that for consistent perturbative expansion we have to impose the value of the Higgs quartic couplings at the electroweak scale to be parametrically given as $\lambda\sim (g^4, g^{\prime^4}, y_t^4)$. This allows us to neglect the $\lambda$-terms in (\ref{central2}) associated with the Higgs and Goldstones and we obtain ($i=W, Z, t$ and neglecting the light masses $m_{light}$):
\begin{eqnarray}
\label{indVC}
&&\mu\frac{\pa \rho_{ind}}{\pa \mu}=\frac{ \langle\phi \rangle^4 }{32\pi^2}\sum_{i} n_i \kappa_i^2\ \int_0^1 \frac{x(1-x)\mu^2 dx}{(M_{phys }^2)_i+x(1-x)\mu^2}\\
&&= \frac{\mu^2}{12 (4\pi)^2}\Bigg[-12 M_t^2+  6 M_{W}^2+ 3 M_Z^2\Bigg] +  \frac{\mu^4}{ 20 (4\pi)^2} \nonumber \\
&& 
%= \frac{\mu^2\langle\phi\rangle^2}{12 (4\pi)^2}\sum_i \kappa_i n_i + \frac{\mu^4}{ 20 (4\pi)^2}
=\frac{\mu^2\langle\phi\rangle^2}{12 (4\pi)^2}\frac{\pa}{\pa \phi^2}\sum_i n_i M_i^2(\phi) +  \frac{\mu^4}{ 20 (4\pi)^2}\,, \nonumber
\end{eqnarray}
where in the last line we used the fact that in the massless theory with only one background field $\phi$, \emph{any} mass can be written as $M_i^2(\phi)=\kappa_i \phi^2$.
%We will now show that the sum in the $\mu^2$-term is identical to Veltman condition in the massless theory.
Notice that the $\mu^2$-proportional term is nothing but the generalization of the well known Veltman condition i.e. the requirement of the absence of the quadratic divergence \footnote{We are working in the Landau gauge, however the 1-loop quadratically
divergent part is gauge invariant in R$_\xi$ gauges \cite{Fukuda:1975di}.} for the Higgs mass (cancellation of the prefactor of the $\phi^2$-term). {\it This means that in the massless case 
%within this class of models 
the fine-tuning problem of the Higgs mass is linked to the fine-tuning problem of the Cosmological Constant value}
%and correspondingly is defined as
%($\mu\approx v= 246$ GeV)
%\be
%V.C.\equiv \frac{\pa}{\pa \phi^2}\sum_i n_i M_i^2(\phi) 
%%\qquad 
%%V.C.^{\rm (SM)}= 3 \lambda(\mu) + \frac{9}{4}g^2(\mu)+\frac{3}{4}g^{\prime 2}(\mu)-6y_t^2(\mu)=0
%%\simeq \frac{1}{\langle\phi \rangle^2}\Bigg[-12 m_t^2+ 6 m_{W}^2+3 m_Z^2\Bigg]=0 \ ,
%\label{VC}
%\ee
\footnote{ In the massive version of the SM, where Higgs mass is generated at tree-level via Higgs mechanism, the Veltman condition leads instead to
\be
V.C.^{\rm (SM)}= 3 \lambda(\mu) + \frac{9}{4}g^2(\mu)+\frac{3}{4}g^{\prime 2}(\mu)-6y_t^2(\mu)=0
%\simeq \frac{1}{\langle\phi \rangle^2}\Bigg[-12 m_t^2+ 6 m_{W}^2+3 m_Z^2\Bigg]=0 \ ,
\label{VCSM}
\ee
and this is a relation among the dimensionless SM couplings at some energy scale $\mu$. From (\ref{massextr}), $M_H^2=\lambda\langle\phi \rangle^2=2m^2$ and using $\mu \approx v_{EW}= 246$ GeV, Eq.(\ref{VCSM}) leads to:
\be
V.C.^{\rm (SM)}= 3 M_H^2 -12 M_t^2+ 6 M_{W}^2+3 M_Z^2=0
%\simeq \frac{1}{\langle\phi \rangle^2}\Bigg[-12 m_t^2+ 6 m_{W}^2+3 m_Z^2\Bigg]=0 \ ,
\label{VC1}
\ee
predicting $M_H\approx 314$ GeV, which is in conflict with experiment too. 
} .
%Now, in the massless limit $m=0$, we have to neglect $m_H$ since, as we discussed above, Higgs mass is generated at one-loop order and after this we recover the $\mu^2$-term in (\ref{indVC})
\subsection{Massless Standard Model with extra massless real scalar}
%The fact that Veltman condition in the massless theory is equivalent  to the requirement that $\mu^2$-term in the $\rho_{ind}$ vanishes is general.
%Indeed, notice that and therefore we can rewrite Veltman condition (\ref{VC}) as:
%\be
%V.C.\equiv\frac{\pa}{\pa \phi^2}\sum_i n_i M_i^2(\phi) =\frac{1}{\phi^2}\sum_i n_i M_i^2(\phi)=\sum_i \kappa_i n_i  \ .
%%\qquad \Longrightarrow \qquad V.C.=\frac{1}{\phi^2}\sum_i n_i M_i^2(\phi)=\sum_i \kappa_i n_i
%\label{VCmassless}
%\ee
%
%
Let us now consider the simplest extension of the SM by adding one extra massless real scalar $S$:
%Let us also assume, for simplicity, that $\langle S \rangle$=0.
%\item  {\bf Massive version:} 
\be
V_0 = V_0^{SM}+ \lambda_{HS} \Phi^\dagger \Phi \, S^2 + \frac{\lambda_S}{4} S^4
%% + \frac{1}{2}m_S S^2\ 
\label{PNC1}
%%%+y_\chi S (\chi \chi +   \bar{\chi} \bar{ \chi}) +  c.t. 
\ee
so that contribution from the Higgs background to the mass of the scalar $S$ is given by $M_S^2(\phi)=\lambda_{HS} \phi^2$.
%\begin{eqnarray}
%&&\mu\frac{\pa (\rLV+\rho_{ind})}{\pa \mu}= {\displaystyle\sum_{j}^{}}{\displaystyle\frac{n_j  (m_{light}^4)_j }{32\pi^2}}+\frac{m^4}{8\pi^2}+\frac{m_S^4}{32\pi^2}-\nonumber \\
%&&\frac{\mu^2}{48 (4\pi^2)}\Bigg[12 m_t^2- 6 m_{W}^2-3 m_Z^2+M_S^2+\frac{m_H^4}{2}\Bigg(\frac{1}{m_W^2}+\frac{1}{2 m_Z^2}-\frac{3}{2m_H^2}\Bigg)\Bigg]
%\label{central3}
%\end{eqnarray}
%
%and therefore, Veltman condition is equivalent to our requirement of absence of leading $\mu^2$- term in (\ref{central2})
%In, the massless SM with addition of SM-singlet massless real scalar $S$ was considered
%\be
%V_0 = V_0^{SM}+ \lambda_{HS} H^\dagger H S^2 + \frac{\lambda_S}{4} S^4 \ 
%\label{PNC}
%%%+y_\chi S (\chi \chi +   \bar{\chi} \bar{ \chi}) +  c.t. 
%\ee
%In the massless theory at the tree-level $\langle\phi \rangle=0$, which means that the tree-level mass of the Higgs is zero and the electroweak symmetry need to be broken radiatevely. For this to happen, we need to balance the tree-level potential against the 1-loop contribution, so that for consistent perturbative expansion we have to impose that the value of the Higgs quartic couplings at the electroweak scale is parametrically given as $\lambda(\mu)\sim (g^4, g^{\prime^4}, y_t^4, \lambda_{HS}^2)$. This allows us to neglect the $\lambda$ in (\ref{VC}) and adding the contribution from $\lambda_{HS}$ we finally obtain
%and using $m_W^2 = v^2 g^2(\mu_0)/4$,  \mbox{$m_Z^2 = v^2(g^2(\mu_0)+{g^\prime}^2(\mu_0))/4$}, $m_t=y_t v/\sqrt{2}$ and $m_S^2 = \lambda_{HS} v^2$
%we notice that this is just Eq.\ref{VC}. 
In this model, the solution to the Veltman condition (\ref{indVC}) reads \footnote{Notice that in (\ref{indVC}) the value of $\mu$ is evaluated at the cosmological RG scale $\mu\approx \mu _{c}$ which is experimentally given by $\mu _{c}=\mathcal{O}(10^{-3})\,eV$. Naively, one would think that we also need to impose the Veltman condition for the SM dimensionless couplings evaluated at this cosmological scale. However, due to decoupling of the SM degrees of freedom below the electroweak scale, the SM beta functions for the dimensionless couplings will scale as $\mu^2/m^2$ so that coupling values "freeze" quickly at their electroweak scale values and therefore the Veltman condition will remain the same up to $\mu^2/m^2$ effects.}  \be
12 M_t^2- 6 M_{W}^2-3 M_Z^2-M_S^2=0 \ \ \Longrightarrow \ \ \lambda_{HS}(\mu) &= 6y_t^2(\mu)-\frac{9}{4}g^2(\mu)-\frac{3}{4}g^{\prime 2}(\mu) \stackrel{\mu\approx v_{EW}}{\approx} 4.8 \ .\nonumber 
%\lambda_S(\mu_0) &= \frac{8}{3} y_\chi^2(\mu_0)-\frac{4}{3} \lambda_{HS}(\mu_0) \stackrel{\mu_0 \approx v}{\approx} \frac{8}{3} y_\chi^2(\mu_0) - 6.45 \ ,
\ee
Working in the parameter space of the model where $\langle S \rangle$=0, see \cite{Antipin:2013exa} for details, leads to the scalar mass $M_S=\sqrt{\lambda_{HS}}\ v_{EW} \approx 550$ GeV which we already noticed above in the massive version of the SM where the role of the scalar $S$ was played by the Higgs. %\footnote{This is the  predicted value for the Higgs mass obtained in  \cite{Babic:2001vv} because from the point of view of counting degrees of freedom in (\ref{IRB}) we just replaced the real Higgs scalar with the new real scalar $S$.}.
%The second solution sets a lower bound on $y_\chi$ from the stability bound on $\lambda_S$, i.e. $y_\chi(\mu_0) \geq 1.55$.%
Remarkably, with the mass of the scalar $S$ satisfying the Veltman condition, we correctly \emph{predict} the one-loop induced Higgs mass from the Coleman-Weinberg potential
\be
M^2_{H} &= \frac{3}{8\pi^2} \big[\frac{1}{16}\big(3g^4+2 g^2 g'^2+g'^4\big)-y_t^4+\frac{1}{3} \lambda_{HS}^2\big] v_{EW}^2  \quad  
&\Longrightarrow \quad M_H \approx 125 \ \text{GeV}\ .
\ee
This provides an interesting example of how the demand for the absence of leading RG effects in the running of the $\rho_{ind}$  due to the heavy particles may provide the hints on the possible extensions of the SM. Moreover, in this model there is no problem with nucleosynthesis for either of the choices for the RG scale $\mu\sim T$ or $\mu\sim H$.

\section{Standard Model in the constant curvature space}

%In our final chapter, 
For our final generalization of (\ref{centralMOM}), we work with the full renormalized version of the Hilbert-Einstein action (\ref{HE})
%with a running cosmological and gravitational
containing additional coupling constants $\kappa=(16\pi G)^{-1}=M_{pl}^2/2$, and non-minimal coupling $\xi$. We consider the Standard Model in the constant curvature space $R_{\mu\nu}=(R/4)g_{\mu\nu}$ and working in the linear curvature approximation in Appendix B we show that appropriate generalization of (\ref{central}) is given by
\be
{\displaystyle \mu \frac{\partial (\rLV+\rho_{ind}+\kappa R)}{\partial \mu}}={\displaystyle\sum_{i}^{}}{\displaystyle\frac{n_i}{32\pi^2}} \mathcal{M}_i^4(\langle\phi \rangle) 
\label{central22}
\ee
where
\be
\mathcal{M}_i^2(\langle\phi \rangle)=\kappa_i\langle\phi \rangle^2-\kappa_i^m m^2 +\bigg(\kappa_i^{R}-\frac{1}{6}\bigg)  R
%\equiv m^2_i(\langle\phi \rangle)+\bigg(\kappa_i^{R}-\frac{1}{6}\bigg)  R
%= \kappa_i \frac{2 (m^2-\xi R)}{\lambda} -m^2\kappa_i^m+\kappa_i^R R  
\ ,
\label{massextr2}
\ee
with parameters $\kappa_i^R$ defined in Table.\ref{SMmasses} in Appendix B.
Generalizing to the mass-dependent scheme we obtain (see Appendix B for details):
\begin{eqnarray}
\label{Rdec}
&&{\displaystyle \mu \frac{\partial (\rLV+\rho_{ind}+\kappa R)}{\partial \mu}}={\displaystyle\sum_{i}^{}}{\displaystyle\frac{n_i}{32\pi^2}} \mathcal{M}_i^4(\langle\phi \rangle) \int_0^1 \frac{x(1-x)\mu^2 dx}{\mathcal{M}^2_i(\langle\phi \rangle)+x(1-x)\mu^2}=\nonumber \\
&&{\displaystyle\sum_{j}^{}}{\displaystyle\frac{n_j  (\mathcal{M}_{light}^4)_j }{32\pi^2}}+\frac{\mu^2}{12 (4\pi)^2}\Bigg[-12 \widetilde{m}_t^2+6 \widetilde{m}_{W}^2+3 \widetilde{m}_Z^2+\widetilde{m}_H^2+\frac{7}{3} R\Bigg]+\frac{\mu^4}{30 (4\pi)^2} \ , 
\end{eqnarray}
where masses $\widetilde{m}_i^2$ have corrections from the non-minimal Higgs coupling $\xi$
\be
\widetilde{m}^2_i\equiv M_i^2(\langle\phi\rangle)-2\kappa_i\frac{ \xi R}{\lambda}\ 
\ee 
with $M_i^2(\langle\phi\rangle)$ defined in (\ref{massextr}). The result (\ref{Rdec}) generalizes effective theory expansion (\ref{betaM}) to the constant curvature space
\begin{equation}
\beta \big(m_{light},\frac{\mu }{m}\big)=a_1\,m_{light}^{4}+b_1\,
\mu^{2}m^{2} +
c_1 \,\mu^{4} + d_1 \mu^{2}R%
+\mathcal{...} 
\label{betaM1}
\end{equation}
which also appears via explicit calculations on the expanding cosmological background where vacuum energy is dynamical \cite{Kohri:2016lsj}. The result (\ref{Rdec}) also generalizes the flat space result to possibility of, for example, curvature-induced running of the vacuum energy and curvature-induced phase transitions \cite{Elizalde:1993ee,Elizalde:1993ew,Elizalde:1993qh,Gorbar:2003yt,Gorbar:2003yp,Markkanen:2014poa}.

%\end{itemize}

\section{Conclusions}
\label{Concl}
 We revisited the decoupling effects associated with heavy particles in the RG running of the vacuum energy using the mass-dependent renormalization scheme. We derived the universal one-loop beta function of the vacuum energy $\rLV+\rho_{ind}$, arising from the Higgs vacuum  and the Cosmological Constant term in the entire energy range, valid in the UV and in the IR regime. 
We have shown that although $\rLV$ and $\rho_{ind}$ run separately, it is only the sum $\rLV+\rho_{ind}$ that exhibits behavior consistent with the decoupling theorem. 

At the energy scale lower than the mass of the particle, the leading term in the RG running of $\rLV+\rho_{ind}$ is proportional to the square of the mass of the heavy particle which leads to the enhanced RG running and, consequently, severe fine-tuning problem with the measured value of the Cosmological Constant. We show that the condition of absence of this leading effect is not satisfied in the SM,  while in the massless theories, where Higgs mass is generated radiatively via Coleman-Weinberg mechanism, this constraint formally coincides with Veltman condition. In a simple extension of the SM with addition of one massless real scalar the  condition of absence of leading effect in $\beta_{\rho_{ind}}$ allowed us to {\it predict} the radiative Higgs mass correctly.

Finally, we also provided the generalization to the constant curvature space in the linear curvature approximation finding the effective field theory expansion that also appears via explicit calculations on the expanding cosmological background. In view of this, our results also might have impact on the models based on the dynamical cosmological constant which were confronted with new cosmological observations in \cite{Sola:2017jbl,Sola:2016zeg} with the results being still inconclusive \cite{Geng:2017apd,Heavens:2017hkr}.

\acknowledgments
We thank F.Sannino, I.Shapiro and H.Stefancic for useful discussions and careful reading of the manuscript. This work is supported by the Croatian Science Foundation (HRZZ) project "Terascale Physics for the LHC and Cosmos" and by the H2020 CSA Twinning project No.692194, RBI-T-WINNING. BM also acknowledges partial support by the Croatian Science Foundation (HRZZ)
project "Physics
of Standard Model and Beyond".

\appendix

%\section{Weak scale thresholds at one loop}\label{1loop}
%We summarise here the one-loop corrections $\theta^{(1)}$
%to the various SM parameters 
%\be
%\theta=\{\lambda, m, y_t, g_2, g_Y\}= 
%\theta^{(0)}+\theta^{(1)}+\theta^{(2)}+\cdots.
%\ee
%We perform one-loop computations in a generic $\xi$ gauge,
%confirming that $\theta^{(1)}$ is gauge-independent, as it should.
%Our expressions for $\theta^{(1)}$ are equivalent to the well known
%expressions in the literature.  We write $\theta^{(1)}$ in terms of
%finite parts of the xBelow we report the expressions valid in the limit $M_b=M_\tau=0$; the negligible effect of
%light fermions masses is included in our full code.

\section{Mass-dependent scheme derivation of the CC running}
In the renormalized perturbation theory, one rewrites the bare parameter $\theta_0$ as
\be
\theta_0 = \theta_{os} - \delta \theta_{os} = \theta_{\overline{MS}} (\mu)- \delta \theta_{\overline{MS}}=
\theta_{\MS}(\mu)-{\delta  \theta}_{\MS}
\label{eq:g1} 
\ee
%or 
%\be
%\theta(\mu) = \theta_{os} - \delta \theta_{os} + 
%{\delta  \theta}_{ \MS}\, ,
%\label{eq:g2} 
%\ee 
where we use on-shell (OS), momentum subtraction (MOM) and ${\overline{MS}}$ schemes with $\theta_{\MS} (\mu)$, $\theta_{os}$, $\theta_{\overline{MS}}$ as
the renormalized MOM, OS, $\overline{MS}$ parameters and 
$\delta  \theta_{\MS }$, $\delta  \theta_{os}$, $\delta \theta_{\overline{MS}}$ as
corresponding counterterms. By definition $\delta  \theta_{\MS }$
subtracts, in the dimensional regularization, the usual $\delta \theta_{\overline{MS}}$ structure
$1/\epsilon+\gamma -\ln (4 \pi)$ and, in addition, makes a finite subtraction of the value of the quantity at $p^2=-\mu^2$.
Concerning  the structure of the $1/\epsilon$ poles
between any two schemes, one notices that it should 
be identical once the poles in one scheme are expressed in terms of 
the quantities of the other scheme. Therefore the difference between the counterterms in two schemes
is finite. Using the fact, that $\theta_{os} $ is physical ($\mu$-independent) parameter, this allows to extract the beta function as, for example,
\be
\theta_{\overline{MS}}(\mu) = \theta_{os} -(\delta \theta_{os}-\delta \theta_{\overline{MS}})|_{\rm fin}\qquad \Longrightarrow \qquad \beta_{\theta_{\overline{MS}}}=-\mu\frac{\pa}{\pa\mu} (\delta \theta_{os}-\delta \theta_{\overline{MS}})|_{\rm fin}.
\label{eq:g3} 
\ee
Now, to calculate the beta function in the MOM scheme, one calculates the derivative $\mu\pa/\pa \mu$ of the renormalized form-factor
ignoring the possible kinematical factors associated with the external momentum. So, our goal is to find the external momentum dependence of the
SM parameters $\theta=(\lambda, m)$.

%Concerning  the structure of the $1/\epsilon$ poles
%in the  OS and MOM counterterms, one  notices that it should 
%be   identical once the poles in the OS counterterms are expressed in terms of 
%the MOM quantities. Then, after this
%operation is performed, the desired $\theta(\mu)$ to 1-loop order is obtained from
%\be
%\theta(\mu) = \theta_{os} - \left. \delta \theta_{}^{(1)} \right|_{\rm fin}^{os-\MS} ,
%\label{eq:g3} 
%\ee
%where the subscript `fin' denotes the finite part of the quantity involved. 
To achieve this, we can use the results of \cite{Buttazzo:2013uya}, where the finite part of $(\delta \theta_{os}-\delta \theta_{\overline{MS}})|_{\rm fin}$
 %$\left. \delta \theta_{}^{(1)} \right|_{\rm fin}^{os-\overline{MS}}$ 
 was provided in terms of the one- and two-point Passarino-Veltman functions:
\be A_0 (M) =
M^2\Bigg(1-\log\frac{M^2}{\mu_{\overline{MS}}^2}\Bigg)\ ,\quad B_0(p;M_1,M_2) = -\int_0^1
\log\frac{xM_1^2+(1-x) M_2^2-x(1-x)p^2}{\mu_{\overline{MS}}^2}dx\ ,  \nonumber\\
\ee
which are connected as 
\begin{eqnarray}
A_0(M) &=& M^2 \left [ B_0(0,M,M) + 1 \right ]\,,
\end{eqnarray}
and
\begin{eqnarray}
\mu\frac{\pa A_0 (M)}{\pa \mu} &=& M^2 \mu\frac{\pa B_0 (0,M,M)}{\pa \mu}
%=M^2 \mu\frac{\pa B_0 (p,M,M)}{\pa \mu} 
\,.
\end{eqnarray}
In the MOM scheme the following relation is valid
\begin{eqnarray}
\mu\frac{\pa A_0 (M)}{\pa \mu}_{|\rm MOM}&=& M^2 \mu\frac{\pa B_0 (0,M,M)}{\pa \mu}_{|\rm MOM}
=M^2 \mu\frac{\pa B_0 (p,M,M)}{\pa \mu}_{|\rm MOM} 
\,.
\end{eqnarray}
which stems from the fact that
\be
B_0(p;M_1,M_2)_{|\rm MOM} = -\int_0^1
\log\frac{xM_1^2+(1-x) M_2^2-x(1-x)p^2}{xM_1^2+(1-x) M_2^2+x(1-x)\mu^2}dx\,.
\ee
Moreover, the unrenormalized form  of $A_0$ 
\be
A_0(M) = -M^2 \left [ \frac{2}{n-4} + \log \left ( \frac{M^2}{4 \pi \mu^2} \right ) + \gamma_E -1 \right ] 
\ee
was used in (\ref{Vacfree2A0}).

With the expressions above we can easily reconstruct the external momentum dependence of the renormalized form-factor we are looking for. We use the one-loop result for the quartic coupling \cite{Buttazzo:2013uya} (notice that all masses are physical):
%The one-loop result for quartic coupling is obtained from (\ref{eq:g3}) where $\lambda_{os}=G_F M_H^2/\sqrt{2}$ and :
\begin{eqnarray}
\label{d1Mh}
(\delta \lambda_{os}-\delta \lambda_{\overline{MS}})|_{\rm fin}&=&
-\frac{2}{(4\pi)^2v_{\rm EW}^4}{\rm Re}\bigg[  3M_t^2(M_H^2-4M_t^2) B_0(M_H;M_t,M_t)+3M_H^2 A_0(M_t)+ \nonumber\\
&&+\frac{1}{4}\left(M_H^4-4 M_H^2 M_Z^2+12 M_Z^4\right)B_0(M_H;M_Z,M_Z) +\frac{M_H^2(7M_W^2-4 M_Z^2)}{2(M_Z^2-M_W^2)}  A_0(M_Z)+   \nonumber  \\
&& +\frac{1}{2}(M_H^4-4M_H^2M_W^2+12M_W^4)B_0(M_H;M_W,M_W)-\frac{3M_H^2 M_W^2}{2(M_H^2-M_W^2)} A_0(M_H)+\nonumber\\
&&+\frac{M_H^2}{2}\left(-11 + \frac{3 M_H^2}{M_H^2-M_W^2} -\frac{3 M_W^2}{M_Z^2 - M_W^2}\right) A_0(M_W) +   \\
&&+\frac{9}{4} M_H^4 B_0(M_H;M_H,M_H) +\frac{1}{4}(M_H^4 +M_H^2(M_Z^2+2M_W^2-6 M_t^2)-8(M_Z^4+2M_W^4))
 \bigg] \ .\nonumber
\end{eqnarray}
Using (\ref{eq:g3}), it is easy to show that (\ref{d1Mh}) leads to:
\begin{eqnarray}
\label{d1Mh1}
\left[{\displaystyle\frac{\beta_{\lambda}}{8}}-{\displaystyle\frac{\gamma_\phi\lambda}{2}}\right]_{\overline{MS}}&=&
\frac{\mu_{\overline{MS}}}{ 2(4\pi)^2v_{\rm EW}^4} \frac{\pa}{\pa \mu_{\overline{MS}}}\bigg[  -12M_t^4 B_0(M_H;M_t,M_t) +\frac{1}{4}\left(M_H^4+12 M_Z^4\right)B_0(M_H;M_Z,M_Z) +   \nonumber  \\
&& +\frac{1}{2}(M_H^4+12M_W^4)B_0(M_H;M_W,M_W)+\frac{9}{4} M_H^4 B_0(M_H;M_H,M_H) 
 \bigg] \ .
\end{eqnarray}

To obtain the corresponding object in the MOM scheme, we only have to reinstate $B_0(M_H;M_1,M_2)\to B_0(p;M_1,M_2)$, make a finite subtraction at $p^2=-\mu^2$ and  calculate the derivative $\mu\pa/\pa \mu$
%$B_0(M_h;M_1,M_2)\to B_0(p;M_1,M_2).$
\footnote{When converting
(\ref{d1Mh}) to the MOM scheme, there will be an additional external momentum dependence due to kinematics. 
%For example, contribution to the Higgs boson self-energy from the W-loop has additional external momentum dependence as :
%\begin{eqnarray}
%OS:&&\quad [\Pi ^{HH}_{(WW)}](p^2=M_h^2)\sim (M_h^4-4M_h^2M_W^2+12M_W^4)\ B_0(M_h;M_W,M_W) \to\\
%MOM:&& \quad [ \Pi ^{HH}_{(WW)}](p^2)\sim  [M_h^4+4M_h^2M_W^2- 4 M_W^2(2p^2-3 M_W^2)] \ B_0(p;M_W,M_W) \ .\nonumber
%\end{eqnarray}
However, as discussed, only momentum dependence of the function $B_0(p;M,M)$ contributes to the beta function in the MOM scheme. 
} .
After that, we arrive at (\ref{scheme}) 
\be
\left[{\displaystyle\frac{\beta_{\lambda}}{8}}-{\displaystyle\frac{\gamma_\phi\lambda}{2}}\right]_{\MS}
=
{\displaystyle\sum_i}{\displaystyle\frac{n_i\kappa_i^2}{32\pi^2}}\int_0^1 \frac{x(1-x)\mu^2 dx}{(M_{phys }^2)_i+x(1-x)\mu^2} \ .
\ee
%where we used the Passarino-Veltman functions 
%The dependence of $\lambda^{(1)}(\mu)$ on the renormalisation scale $\mu$ reproduces
%the well known one-loop RGE equation for $\lambda$.
Similarly, the Higgs mass term is corrected as \cite{Buttazzo:2013uya}:
 \begin{eqnarray}
(\delta m^2_{os}-\delta m^2_{\overline{MS}})|_{\rm fin} &=&
-\frac{1}{(4\pi)^2 v_{\rm EW}^2} {\rm Re} \bigg[  6M_t^2(M_H^2-4M_t^2)B_0(M_H;M_t,M_t)+ 24 M_t^2 A_0(M_t) +\nonumber\\
&& +(M_H^4-4M_H^2M_W^2+12M_W^4)B_0(M_H;M_W,M_W)-2(M_H^2+ 6M_W^2) A_0(M_W) +  \nonumber  \\
&&+\frac{1}{2}\left(M_H^4-4 M_H^2 M_Z^2+12 M_Z^4\right)B_0(M_H;M_Z,M_Z) -(M_H^2+ 6M_Z^2) A_0(M_Z) +  \nonumber  \\
&&+\frac{9}{2} M_h^4 B_0(M_H;M_H,M_H)-3M_H^2 A_0(M_H) \bigg] \ , 
\end{eqnarray}
%and it is easy to show that:
%\begin{eqnarray}
%\label{d1Mh2}
%\left[{\displaystyle\frac{1}{2}}\gamma_{m}-\gamma_\phi\right]_{\overline{MS}}&=&
%-\frac{\mu_{\overline{MS}}}{(4\pi)^2V^4} \frac{\pa}{\pa \mu_{\overline{MS}}}\bigg[ \frac{1}{2}M_h^4B_0(M_h;M_Z,M_Z) +   \nonumber  \\
%&& +M_h^4B_0(M_h;M_W,M_W)+\frac{3}{2} M_h^4 B_0(M_h;M_h,M_h) 
% \bigg] \ .
%\end{eqnarray}
and repeating the same steps as in the $\lambda$ case above, we obtain (\ref{schemem}):
\be
\left[{\displaystyle\frac{1}{2}}\gamma_{m}-\gamma_\phi\right]_{\MS}=
{\displaystyle\sum_i}{\displaystyle\frac{n_i\kappa_i\kappa_i^{m}}{16\pi^2}} \int_0^1 \frac{x(1-x)\mu^2 dx}{(M_{phys }^2)_i+x(1-x)\mu^2} \,.
\label{schemem1}
\ee
The $\beta_{\lambda}$, $\beta_{m^2}$ and $\gamma_\phi$ are one-loop
$\beta$- and $\gamma$-functions:
\begin{eqnarray}
\label{betas}
16\pi^2\beta_{\lambda}&=&12\left(\lambda^2-y^4_t+\lambda y_t^2\right) -
\left(3g'^2+9g^2\right) \lambda
+{\displaystyle\frac{9}{4}}\left[{\displaystyle\frac{1}{3}}g'^4
+{\displaystyle\frac{2}{3}}g'^2g^2+g^4\right] \ ,
\vspace{0.2cm}\nonumber\\
16\pi^2\beta_{m^2}&=&16\pi^2 \gamma_{m} m^2=m^2\left[6\lambda+6
y_t^2-{\displaystyle\frac{9}{2}}g^2-
{\displaystyle\frac{3}{2}}g'^2\right] \ ,
\vspace{0.2cm}\nonumber\\
16\pi^2\gamma_\phi&=&3\,\left(y^2_t-
{\displaystyle\frac{1}{4}}g'^2-
{\displaystyle\frac{3}{4}}g^2\right) \ .
\end{eqnarray}

\section{Generalization to constant curvature space}
We work in the constant curvature space $R_{\mu\nu}=(R/4)g_{\mu\nu}$ and in the linear curvature approximation we consider 
%(for convenience, all the differences with the flat spacetime ($R=0$) calculation above are highlighted in \textcolor{red}{red}):
\be
\label{veff1}
V(\rLV,\phi, m^2,\lambda_i,\kappa,\xi,\mu) \equiv V_0 + V_1 + \cdots \;\; ,
\ee
where $\lambda_i\equiv (g,g',\lambda,h_t)$ runs over
SM dimensionless couplings and $V_0$, $V_1$ are respectively
the tree level potential and the one-loop correction, namely \cite{Barvinsky:2009fy}
%\subequations{
\be
\label{v01}
V_0=-{\displaystyle\frac{1}{2}}m^2\phi^2 +
{\displaystyle\frac{1}{8}}\lambda\phi^4 +\frac{1}{2}\xi R \phi^2 \ ,
\ee
\be
\label{v11}
V_1={\displaystyle\sum_{i}}{\displaystyle\frac{n_i}{64\pi^2}}
\bigg(\bar{M}_i^4(\phi)-\frac{\bar{M}_i^2(\phi)R}{3}\bigg)\log{\displaystyle\frac{\bar{M}_i^2(\phi)}{\mu^2_{\overline{MS}}}}
+\rLV \ + \kappa R  \ ,
\ee
%}
%\endsubequations
where we showed only logarithmic term relevant for us and defined
\be
\bar{M}_i^2(\phi)=\kappa_i\phi^2-\kappa_i^m m^2 + \kappa_i^R R \ ,
\label{Rmasses}
\ee
with the parameters $n_i$, $\kappa_i$, $\kappa_i^m$ and $\kappa_i^R$ shown in Table.\ref{SMmasses}.
%\begin{center}
\begin{table}
\center
\caption{\label{SMmasses}Contributions to the effective potential (\ref{v11})
 from the SM particles $W^{\pm}$, $Z^0$, top quark t, Higgs $\phi$ and the Goldstone bosons $\chi_{1,2,3}$ \cite{Moss:2015gua}.}
%\vspace{2mm}
 \begin{tabular}{|c||ccccc|}
 \hline
   $\Phi$ & $~~i$ & $~~n_i$  &$~~\kappa_i$ & $\kappa_i^m$          & $~~\kappa_i^R$    \\[1mm]\hline
   $~~W^\pm(\rm ghost)$ & $~~1$  & $-2$       & $~~ g^2/4$        & $0$        & $~~{1}/{2}$     \\[1mm]
   $~W^\pm$ & $~~2$  & $~~8$       & $~~ g^2/4$        & $0$        & $~~{1}/{2}$ \\[1mm]      \hline
   $Z^0(\rm ghost)$ & $~~3$  & $-1$       & $~~(g^2+g'^2)/4$ & $0$        & $~~{1}/{2}$      \\[1mm]
   $Z^0$ & $~~4$  & $~~4$       & $~~(g^2+g'^2)/4$ & $0$        & $~~{1}/{2}$      \\[1mm] \hline
   t & $~~5$  & $-12$     & $~~ y_{\rm t}^2/2$      & $0$        & $~~{1}/{4}$      \\[1mm]\hline
   $\phi$ & $~~6$  & $~~1$       & $~~3\lambda/2$           & $~1$      & $~~1/2~$ 
\\[1mm]\hline
   $\chi_i$ & $~~7$  & $~~3$       & $~~\lambda/2$            & $~1$      & $~~1/2$   
\\[.5mm]\hline
  \end{tabular}
  \end{table}
$\bar{M}_i^2(\phi)$ are the tree-level expressions for the background-dependent and curvature-dependent masses of the
particles that enter in the one-loop radiative corrections.
 Also $\kappa=(16\pi G)^{-1}=M_{pl}^2/2$ and $\xi$ is the non-minimal coupling.
It is convenient to redefine $\mathcal{M}_i^2\equiv\bar{M}_i^2-R/6$ so that (up to $R^2$-terms)
\be
\label{v11}
V_1={\displaystyle\sum_{i}}{\displaystyle\frac{n_i}{64\pi^2}}
\mathcal{M}_i^4(\phi)\log{\displaystyle\frac{\mathcal{M}_i^2(\phi)}{\mu^2_{\overline{MS}}}}
+\rLV \ + \kappa R  \ .
\ee

We again split the potential to vacuum and $\phi$-dependent pieces
%$\phi$-dependent "scalar" term:
\be
V(\rLV,\phi,m^2,\la_i,\kappa,\xi,\mu)=
V_{vac}(\rLV,m^2,\la_i,\kappa,\xi,\mu) +V_{scal}(\phi,m^2,\la_ i,\xi,\mu)  \ .
\ee
and the RG equations (\ref{RG0} - \ref{RG-sep0}) get now modified as follows:
\be 
\Big(\mu\frac{\pa}{\pa \mu}+\beta_{\la_i}\frac{\pa}{\pa \la_i} +\ga_m m^2
\frac{\pa}{\pa m^2} -\ga_\phi \phi \frac{\pa}{\pa \phi} +\beta_{\rLV}\frac{\pa}{\pa \rLV}+\beta_{\kappa}\frac{\pa}{\pa \kappa}+\beta_{\xi}\frac{\pa}{\pa \xi}\Big)V(\rLV,\phi, m^2,\lambda_i,\kappa,\xi,\mu)\,=\,0\, \label{RG1}\nonumber \\
\ee
%From the RG-invariance of the renormalized EA -- see
%Eq.\,(\ref{nn8}) -- it follows immediately the $\mu$-independence of
%the renormalized functional $\Ga_{vac}$ and, therefore, we arrive at
%the second identity (\ref{RG-sep}) for the vacuum part of the
%effective potential, while the first identity is the result of the
%subtraction of (\ref{RG-sep}) from (\ref{RG}). The net result is
%that the vacuum and matter parts of the effective potential are
%overall $\mu$-independent separately and no cancelation between them
%can be expected. \vskip 1mm
%Using \ref{Veff}, we now show that Eq.\,(\ref{RG}) is, in fact, a sum of two independent RG equations,
%
\begin{eqnarray}
&&\Big(\mu\frac{\pa}{\pa \mu} + \beta_{\la_i} \frac{\pa}{\pa \la_i}+\ga_m
m^2 \frac{\pa}{\pa m^2}+\beta_{\rLV}\frac{\pa}{\pa \rLV}+\beta_{\kappa}\frac{\pa}{\pa\kappa}+\beta_{\xi}\frac{\pa}{\pa \xi}\Big)V_{vac}(\rLV,m^2,\la_i,\kappa,\xi,\mu) \,=\,0\,,\nonumber\\
\label{RG-sep1}
\\
&&\Big(\mu\frac{\pa}{\pa \mu} + \beta_{\la_i}\frac{\pa}{\pa \la_i}
+\ga_m m^2 \frac{\pa}{\pa m^2} -\ga_\phi \phi \frac{\pa}{\pa \phi}+\beta_{\xi}\frac{\pa}{\pa \xi}
\Big)V_{scal}(\phi,m^2,\la_i,\xi,\mu)\,=\,0\,.
\label{RG-scalar1}
\end{eqnarray}
These equations are valid for any value of $\phi$. However, for the extremum value $\phi=\langle\phi \rangle$ defined via 
 $\frac{\partial V_{scal}(\phi)}{\partial \phi}\Big |_{\langle\phi \rangle}=0$,  the term containing anomalous dimension of the Higgs $\gamma_\phi$ will drop out and we have:
\begin{eqnarray}
&&\Big(\mu\frac{\pa}{\pa \mu} + \beta_{\la_i}\frac{\pa}{\pa \la_i}
+\ga_m m^2 \frac{\pa}{\pa m^2} +\beta_{\xi}\frac{\pa}{\pa \xi}
\Big)V_{scal}(\langle\phi \rangle,m^2,\la_i,\xi,\mu)\,=\,0\, .
\label{RG-scalar11}
\end{eqnarray}

Using the tree-level potential (\ref{v01}), it is useful to define parameter $\rho_{ind}(\mu)=V_0 (\langle\phi \rangle)=-\frac{(m^2(\mu)-\xi(\mu) R)^2}{2\lambda(\mu)}$ which will play the role similar to the $\rLV(\mu)$. The running of this parameter reads:
\be
\mu\frac{\pa \rho_{ind}(\mu)}{\pa \mu}=- \frac{m^2-\xi R}{\la}(\gamma_m m^2 -\beta_\xi R)+\frac{(m^2-\xi R)^2}{2\la^2}\beta_{\la} \ .
%\rho_{ind}(\mu) \  \Big(2\gamma_m -\frac{\beta_{\la}}{\la} +\textcolor{red}{\frac{\beta_{\xi}}{\xi}}\Big) +\textcolor{red}{\frac{m^2 \xi R \gamma_m}{\lambda}} \ .
\label{aux}
\ee
%up to the terms of $\mathcal{O}(R^2)$.

 %$\Lambda_{ind}\equiv V_{scal}(\langle\phi \rangle,m^2,\la_i,\rLV,\mu)$

%\cite{Einhorn},
%which will turn to be irrelevant in our calculation.
By equating the terms with $R$ and different powers of $\phi$ : 
\be
\mu\frac{dV}{d \mu}\sim \big( \ \phi^4 [...]+ m^2 \phi^2[....]+ m^4[...]+m^2 R[... ]+\phi^2 R[... ] \big)=0 \ ,
\ee
 it is straightforward to check that the requirement (\ref{RG1})
applied to the full one-loop effective potential
leads to 
%\subequations{
\begin{align}
\frac{1}{8}\beta_{\lambda}-\frac{1}{2}
\gamma_\phi\lambda=
\sum_i\frac{n_i\kappa_i^2}{32\pi^2} \ ,
\label{p4inv1}
%\vspace{0.3cm}
\end{align}
\begin{align}
{\displaystyle\frac{1}{2}}\gamma_{m}-\gamma_\phi=
{\displaystyle\sum_i}{\displaystyle\frac{n_i\kappa_i\kappa_i^{m}}{16\pi^2}} \ ,
\label{p2inv1}
%\vspace{0.3cm}
\end{align}
\begin{align}
{\displaystyle \mu \frac{\partial \rLV}{\partial \mu}}=m^4 {\displaystyle\sum_i}{\displaystyle\frac{n_i(\kappa_i^{m})^2}{32\pi^2}}
\label{p0inv1} \ ,
\end{align}
\begin{align}
{\displaystyle \mu \frac{\partial \kappa}{\partial \mu}}=m^2{\displaystyle\sum_i}{\displaystyle\frac{n_i \kappa_i^{m}}{16\pi^2 }} \bigg(\kappa_i^{R}-\frac{1}{6}\bigg) 
\label{p0inv0} \ ,
\end{align}
\begin{align}
{\displaystyle\frac{1}{2}}\beta_{\xi}-\xi\gamma_\phi=
{\displaystyle\sum_i}{\displaystyle\frac{n_i\kappa_i}{16\pi^2 }} \bigg(\kappa_i^{R}-\frac{1}{6}\bigg) \ ,
\label{p2inv3}
\end{align}
%}
where SM $\beta$- and $\gamma$-functions are given above in (\ref{betas}), while 
\begin{align}
& 16\pi^2\beta_{\kappa} = \frac{4}{3} m^2  \,, \\
& 16\pi^2(\beta_{\xi}-2\xi \gamma_\phi) = 2\lambda -y_t^2+\frac{g^{\prime 2}}{2}+\frac{3g^2}{2}\,,
\end{align}
with $\beta_{\xi}$ as reported in \cite{Moss:2015gua}.
%For the extremum value $\phi=\langle\phi \rangle$ we have to drop the $\gamma_\phi$ terms from (\ref{p4inv1}),  (\ref{p2inv1}) and (\ref{p2inv3}) and 
Working to linear order in $R$ and using (\ref{p4inv1}),  (\ref{p2inv1}) and (\ref{p2inv3}) in order to reconstruct \ref{aux},
%subtracting these equations appropriately we obtain the running of the $\rho_{ind}$:
%\be
%-\frac{m^4}{2\lambda}\Bigg({\displaystyle 2\gamma_{m}}-{\displaystyle\frac{\beta_{\lambda}}{\la}}\Bigg)={\displaystyle \mu \frac{\partial \rho_{ind}}{\partial \mu}}=
%m^4\Bigg({\displaystyle\sum_i}{\displaystyle\frac{n_i\kappa_i^2}{8\pi^2\la^2}} - {\displaystyle\sum_i}{\displaystyle\frac{n_i\kappa_i\kappa_i^{m}}{8\pi^2m^2\lambda}} \Bigg )
%\label{ind1}
%\ee
%(\ref{}) and (\ref{}) 
we finally obtain
%(remember, $\kappa_4^m=\kappa_5^m=m^2$ ):
\be
{\displaystyle \mu \frac{\partial (\rLV+\rho_{ind}+\kappa R)}{\partial \mu}}={\displaystyle\sum_{i}^{}}{\displaystyle\frac{n_i}{32\pi^2}} \mathcal{M}_i^4(\langle\phi \rangle) 
\label{central1}
\ee
where we used 
\be
\mathcal{M}_i^2(\langle\phi \rangle)=\kappa_i\langle\phi \rangle^2-\kappa_i^m m^2+\bigg(\kappa_i^{R}-\frac{1}{6}\bigg)  R\equiv \widetilde{m}^2_i(\langle\phi \rangle)+\bigg(\kappa_i^{R}-\frac{1}{6}\bigg)  R
%= \kappa_i \frac{2 (m^2-\xi R)}{\lambda} -m^2\kappa_i^m+\kappa_i^R R  
\ ,
\label{massextr1}
\ee
with $\langle\phi \rangle^2=\frac{2 (m^2-\xi R)}{\lambda}$. Notice that for the Goldstones $\mathcal{M}_{\chi }^2(\langle\phi \rangle)\sim R$ and therefore $\mathcal{M}_\chi^4(\langle\phi \rangle)\sim R^2$ so that to linear $R$ order they do not contribute to the sum. The masses $\widetilde{m}^2_i$ formally look identical to the flat space analogues $M_i^2(\langle\phi\rangle)$ in (\ref{massextr}) but, however, contain the curvature corrections via the $\langle\phi \rangle$ :
\be
\widetilde{m}^2_i(\langle\phi \rangle)=\kappa_i\langle\phi \rangle^2 - \kappa_i^m m^2=\kappa_i\frac{2 (m^2-\xi R)}{\lambda}-\kappa_i^m m^2 = M_i^2(\langle\phi\rangle)-2\kappa_i\frac{ \xi R}{\lambda}\ .
\ee 

Generalizing to the mass-dependent scheme we obtain:
\be
{\displaystyle \mu \frac{\partial (\rLV+\rho_{ind}+\kappa R)}{\partial \mu}}={\displaystyle\sum_{i}^{}}{\displaystyle\frac{n_i}{32\pi^2}} \mathcal{M}_i^4(\langle\phi \rangle) \int_0^1 \frac{x(1-x)\mu^2 dx}{\mathcal{M}^2_i(\langle\phi \rangle)+x(1-x)\mu^2} \ .
\label{decR}
\ee
Since the mass of the ghost is equal to the corresponding vector boson mass, the ghost cancels the unphysical gauge mode (see Table.\ref{SMmasses}) and we obtain:
\be
&&{\displaystyle \mu \frac{\partial (\rLV+\rho_{ind}+\kappa R)}{\partial \mu}}={\displaystyle\sum_{j}^{}}{\displaystyle\frac{n_j  (\mathcal{M}_{light}^4)_j }{32\pi^2}}+\frac{\mu^2}{12 (4\pi)^2}\sum_i n_i \mathcal{M}_i^2(\phi)-\frac{\mu^4}{60 (4\pi)^2}\sum_i n_i=\nonumber \\
&&={\displaystyle\sum_{j}^{}}{\displaystyle\frac{n_j  (\mathcal{M}_{light}^4)_j }{32\pi^2}}
%+\frac{\mu^2}{12 (4\pi)^2}\bigg[\frac{7}{3} R+\sum_i n_i m_i^2(\phi)\Bigg]+\frac{\mu^4}{30 (4\pi)^2}=
+\frac{\mu^2}{12 (4\pi)^2}\Bigg[-12 \widetilde{m}_t^2+6 \widetilde{m}_{W}^2+3 \widetilde{m}_Z^2+\widetilde{m}_H^2+\frac{7}{3} R\Bigg]+\frac{\mu^4}{30 (4\pi)^2} \ . 
\ee


\begin{thebibliography}{99}

%\cite{Shapiro:2009dh}
\bibitem{Shapiro:2009dh}
  I.~L.~Shapiro and J.~Sola,
  ``On the possible running of the cosmological 'constant',''
  Phys.\ Lett.\ B {\bf 682} (2009) 105
%  doi:10.1016/j.physletb.2009.10.073
  [arXiv:0910.4925 [hep-th]].
  %%CITATION = doi:10.1016/j.physletb.2009.10.073;%%
  %130 citations counted in INSPIRE as of 14 Feb 2017

%\cite{Shapiro:2008sf}
\bibitem{Shapiro:2008sf}
  I.~L.~Shapiro,
  ``Effective Action of Vacuum: Semiclassical Approach,''
  Class.\ Quant.\ Grav.\  {\bf 25} (2008) 103001
%  doi:10.1088/0264-9381/25/10/103001
  [arXiv:0801.0216 [gr-qc]].
  %%CITATION = doi:10.1088/0264-9381/25/10/103001;%%
  %99 citations counted in INSPIRE as of 27 Mar 2017


 \bibitem{BrownBook}
 L. Brown,
Quantum Field Theory
(Cambridge University Press, Cambridge, 1994).

%\cite{Appelquist:1974tg}
\bibitem{Appelquist:1974tg}
  T.~Appelquist and J.~Carazzone,
  "Infrared Singularities and Massive Fields,''
  Phys.\ Rev.\ D {\bf 11} (1975) 2856.
%  doi:10.1103/PhysRevD.11.2856
  %%CITATION = doi:10.1103/PhysRevD.11.2856;%%
  %1572 citations counted in INSPIRE as of 31 Mar 2017

  %\cite{Casas:1994qy}
\bibitem{Casas:1994qy}
  J.~A.~Casas, J.~R.~Espinosa and M.~Quiros,
  ``Improved Higgs mass stability bound in the standard model and implications for supersymmetry,''
  Phys.\ Lett.\ B {\bf 342} (1995) 171
%  doi:10.1016/0370-2693(94)01404-Z
  [hep-ph/9409458].
%\cite{Casas:1994us}

  %\cite{Ford:1992mv}
\bibitem{Ford:1992mv}
  C.~Ford, D.~R.~T.~Jones, P.~W.~Stephenson and M.~B.~Einhorn,
  ``The Effective potential and the renormalization group,''
  Nucl.\ Phys.\ B {\bf 395} (1993) 17
%  doi:10.1016/0550-3213(93)90206-5
  [hep-lat/9210033].
  %%CITATION = doi:10.1016/0550-3213(93)90206-5;%%
  %306 citations counted in INSPIRE as of 02 Feb 2017
  
 
 \bibitem{Casas:1994us}
  J.~A.~Casas, J.~R.~Espinosa, M.~Quiros and A.~Riotto,
  ``The Lightest Higgs boson mass in the minimal supersymmetric standard model,''
  Nucl.\ Phys.\ B {\bf 436} (1995) 3
   Erratum: [Nucl.\ Phys.\ B {\bf 439} (1995) 466]
 % doi:10.1016/0550-3213(94)00508-C, 10.1016/0550-3213(95)00057-Y
  [hep-ph/9407389].
  %%CITATION = doi:10.1016/0550-3213(94)00508-C, 10.1016/0550-3213(95)00057-Y;%%
    %411 citations counted in INSPIRE as of 25 Feb 2017
    
 %\cite{Shapiro:2008yu}
\bibitem{Shapiro:2008yu}
  I.~L.~Shapiro and J.~Sola,
  ``Can the cosmological 'constant' run? - It may run,''
  arXiv:0808.0315 [hep-th].
  %%CITATION = ARXIV:0808.0315;%%
  %43 citations counted in INSPIRE as of 14 Feb 2017
  
\bibitem{Spencer-Smith:2014woa}
  A.~Spencer-Smith,
  ``Higgs Vacuum Stability in a Mass-Dependent Renormalisation Scheme,''
  arXiv:1405.1975 [hep-ph].
  %%CITATION = ARXIV:1405.1975;%%
  %43 citations counted in INSPIRE as of 12 Mar 2017  
  
  %\cite{Gorbar:2002pw}
\bibitem{Gorbar:2002pw}
  E.~V.~Gorbar and I.~L.~Shapiro,
  ``Renormalization group and decoupling in curved space,''
  JHEP {\bf 0302} (2003) 021
%  doi:10.1088/1126-6708/2003/02/021
  [hep-ph/0210388]; 

%\cite{Babic:2001vv}
\bibitem{Babic:2001vv}
  A.~Babic, B.~Guberina, R.~Horvat and H.~Stefancic,
  ``Renormalization group running of the cosmological constant and its implication for the Higgs boson mass in the standard model,''
  Phys.\ Rev.\ D {\bf 65} (2002) 085002
%  doi:10.1103/PhysRevD.65.085002
  [hep-ph/0111207].
  %%CITATION = doi:10.1103/PhysRevD.65.085002;%%
  %96 citations counted in INSPIRE as of 21 Jan 2017
  
  \bibitem{Shapiro:2000dz}
  I.~L.~Shapiro and J.~Sola,
  ``Scaling behavior of the cosmological constant: Interface between quantum field theory and cosmology,''
  JHEP {\bf 0202} (2002) 006
%  doi:10.1088/1126-6708/2002/02/006
  [hep-th/0012227].

  %\cite{Fukuda:1975di}
\bibitem{Fukuda:1975di}
  R.~Fukuda and T.~Kugo,
  ``Gauge Invariance in the Effective Action and Potential,''
  Phys.\ Rev.\ D {\bf 13} (1976) 3469.
%  doi:10.1103/PhysRevD.13.3469
  %%CITATION = doi:10.1103/PhysRevD.13.3469;%%
  %131 citations counted in INSPIRE as of 25 Feb 2017

  
%  %\cite{Gies:2014xha}
%\bibitem{Gies:2014xha}
%  H.~Gies and R.~Sondenheimer,
%  %``Higgs Mass Bounds from Renormalization Flow for a Higgs-top-bottom model,''
%  Eur.\ Phys.\ J.\ C {\bf 75} (2015) no.2,  68
%  doi:10.1140/epjc/s10052-015-3284-1
%  [arXiv:1407.8124 [hep-ph]].
  %%CITATION = doi:10.1140/epjc/s10052-015-3284-1;%%
  %19 citations counted in INSPIRE as of 03 Feb 2017

  %%CITATION = doi:10.1016/0370-2693(94)01404-Z;%%
  %419 citations counted in INSPIRE as of 29 Jan 2017
  
  %\cite{Antipin:2013exa}
\bibitem{Antipin:2013exa}
  O.~Antipin, M.~Mojaza and F.~Sannino,
  ``Conformal Extensions of the Standard Model with Veltman Conditions,''
  Phys.\ Rev.\ D {\bf 89} (2014) no.8,  085015
%  doi:10.1103/PhysRevD.89.085015
  [arXiv:1310.0957 [hep-ph]].
  %%CITATION = doi:10.1103/PhysRevD.89.085015;%%
  %42 citations counted in INSPIRE as of 21 Jan 2017
  
  \bibitem{Kohri:2016lsj}
  K.~Kohri and H.~Matsui,
  ``Running Cosmological Constant and Renormalized Vacuum Energy Density in Curved Background,''
  arXiv:1612.08818 [hep-th].
  %%CITATION = ARXIV:1612.08818;%% 
  
  %\cite{Elizalde:1993ee}
\bibitem{Elizalde:1993ee}
  E.~Elizalde and S.~D.~Odintsov,
  ``Renormalization group improved effective potential for gauge theories in curved space-time,''
  Phys.\ Lett.\ B {\bf 303} (1993) 240
   [Russ.\ Phys.\ J.\  {\bf 37} (1994) 25]
%  doi:10.1016/0370-2693(93)91427-O
  [hep-th/9302074].
  %%CITATION = doi:10.1016/0370-2693(93)91427-O;%%
  %56 citations counted in INSPIRE as of 10 Apr 2017
  
  %\cite{Elizalde:1993ew}
\bibitem{Elizalde:1993ew}
  E.~Elizalde and S.~D.~Odintsov,
  ``Renormalization group improved effective Lagrangian for interacting theories in curved space-time,''
  Phys.\ Lett.\ B {\bf 321} (1994) 199
%  doi:10.1016/0370-2693(94)90464-2
  [hep-th/9311087].
  %%CITATION = doi:10.1016/0370-2693(94)90464-2;%%
  %49 citations counted in INSPIRE as of 10 Apr 2017
  
  %\cite{Elizalde:1993qh}
\bibitem{Elizalde:1993qh}
  E.~Elizalde and S.~D.~Odintsov,
  %
``Renormalization group improved effective potential for interacting theories with several mass scales in curved space-time,''
  Z.\ Phys.\ C {\bf 64} (1994) 699
%  doi:10.1007/BF01957780
  [hep-th/9401057].
  %%CITATION = doi:10.1007/BF01957780;%%
  %24 citations counted in INSPIRE as of 10 Apr 2017
 



\bibitem{Gorbar:2003yt}
  E.~V.~Gorbar and I.~L.~Shapiro,
  ``Renormalization group and decoupling in curved space. 2. The Standard model and beyond,''
  JHEP {\bf 0306} (2003) 004
%  doi:10.1088/1126-6708/2003/06/004
  [hep-ph/0303124].
  

%\cite{Gorbar:2003yp}
\bibitem{Gorbar:2003yp}
  E.~V.~Gorbar and I.~L.~Shapiro,
  ``Renormalization group and decoupling in curved space. 3. The Case of spontaneous symmetry breaking,''
  JHEP {\bf 0402} (2004) 060
%  doi:10.1088/1126-6708/2004/02/060
  [hep-ph/0311190].
  %%CITATION = doi:10.1088/1126-6708/2004/02/060;%%
  %19 citations counted in INSPIRE as of 20 Mar 2017

\bibitem{Markkanen:2014poa} 
  T.~Markkanen,
  ``Curvature induced running of the cosmological constant,''
  Phys.\ Rev.\ D {\bf 91}, no. 12, 124011 (2015)
 % doi:10.1103/PhysRevD.91.124011
  [arXiv:1412.3991 [gr-qc]].
  
 %\cite{Sola:2016zeg}
\bibitem{Sola:2016zeg}
  J.~Sola,
  ``Cosmological constant vis-a-vis dynamical vacuum: bold challenging the $\Lambda$CDM,''
  Int.\ J.\ Mod.\ Phys.\ A {\bf 31} (2016) no.23,  1630035
%  doi:10.1142/S0217751X16300350
  [arXiv:1612.02449 [astro-ph.CO]].
  %%CITATION = doi:10.1142/S0217751X16300350;%%
  %7 citations counted in INSPIRE as of 31 Mar 2017


%\cite{Sola:2017jbl}
\bibitem{Sola:2017jbl}
  J.~Sola, J.~d.~C.~Perez and A.~Gomez-Valent,
  ``Towards the firsts compelling signs of vacuum dynamics in modern cosmological observations,''
  arXiv:1703.08218 [astro-ph.CO].
  %%CITATION = ARXIV:1703.08218;%%
  %1 citations counted in INSPIRE as of 31 Mar 2017
  
  %\cite{Geng:2017apd}
\bibitem{Geng:2017apd}
  C.~Q.~Geng, C.~C.~Lee and L.~Yin,
  ``Constraints on running vacuum model with $H(z)$ and $f \sigma_8$,''
  arXiv:1704.02136 [astro-ph.CO].
  %%CITATION = ARXIV:1704.02136;%%
  %2 citations counted in INSPIRE as of 11 Aug 2017
  
  %\cite{Heavens:2017hkr}
\bibitem{Heavens:2017hkr}
  A.~Heavens, Y.~Fantaye, E.~Sellentin, H.~Eggers, Z.~Hosenie, S.~Kroon and A.~Mootoovaloo,
  ``No evidence for extensions to the standard cosmological model,''
  arXiv:1704.03467 [astro-ph.CO].
  %%CITATION = ARXIV:1704.03467;%%
  %7 citations counted in INSPIRE as of 11 Aug 2017
   
  
  %\cite{Buttazzo:2013uya}
\bibitem{Buttazzo:2013uya}
  D.~Buttazzo, G.~Degrassi, P.~P.~Giardino, G.~F.~Giudice, F.~Sala, A.~Salvio and A.~Strumia,
  ``Investigating the near-criticality of the Higgs boson,''
  JHEP {\bf 1312} (2013) 089
%  doi:10.1007/JHEP12(2013)089
  [arXiv:1307.3536 [hep-ph]].
  %%CITATION = doi:10.1007/JHEP12(2013)089;%%
  %573 citations counted in INSPIRE as of 11 Mar 2017
  
%%\cite{Kohri:2016lsj}
%\bibitem{Kohri:2016lsj}
%  K.~Kohri and H.~Matsui,
%  %``Running Cosmological Constant and Renormalized Vacuum Energy Density in Curved Background,''
%  arXiv:1612.08818 [hep-th].
%  %%CITATION = ARXIV:1612.08818;%%
%  
  
%  %\cite{Einhorn:1992um}
%\bibitem{Einhorn:1992um}
%  M.~B.~Einhorn and D.~R.~T.~Jones,
%  %``The Effective potential and quadratic divergences,''
%  Phys.\ Rev.\ D {\bf 46} (1992) 5206.
%  doi:10.1103/PhysRevD.46.5206
%  %%CITATION = doi:10.1103/PhysRevD.46.5206;%%
%  %57 citations counted in INSPIRE as of 21 Jan 2017


  
%\cite{Barvinsky:2009fy}
\bibitem{Barvinsky:2009fy}
  A.~O.~Barvinsky, A.~Y.~Kamenshchik, C.~Kiefer, A.~A.~Starobinsky and C.~Steinwachs,
  ``Asymptotic freedom in inflationary cosmology with a non-minimally coupled Higgs field,''
  JCAP {\bf 0912} (2009) 003
%  doi:10.1088/1475-7516/2009/12/003
  [arXiv:0904.1698 [hep-ph]].
  %%CITATION = doi:10.1088/1475-7516/2009/12/003;%%
  %171 citations counted in INSPIRE as of 20 Mar 2017

%\cite{Moss:2015gua}
\bibitem{Moss:2015gua}
  I.~G.~Moss,
  ``Vacuum stability and the scaling behaviour of the Higgs-curvature coupling,''
  arXiv:1509.03554 [hep-th].
  %%CITATION = ARXIV:1509.03554;%%
  %12 citations counted in INSPIRE as of 20 Mar 2017



%\bibitem{Herranen:2014cua} 
 % M.~Herranen, T.~Markkanen, S.~Nurmi and A.~Rajantie,
  %``Spacetime curvature and the Higgs stability during inflation,''
  %Phys.\ Rev.\ Lett.\  {\bf 113}, no. 21, 211102 (2014)
%  doi:10.1103/PhysRevLett.113.211102
  %[arXiv:1407.3141 [hep-ph]].
  
 

 
 

 
 %\bibitem{Garbrecht:2006aw}
 % B.~Garbrecht,
  %``Radiative Lifting of Flat Directions of the MSSM in de Sitter Background,''
  %Nucl.\ Phys.\ B {\bf 784} (2007) 118
%  doi:10.1016/j.nuclphysb.2007.06.013
  %[hep-ph/0612011].
  
\end{thebibliography}
\end{document}